\documentclass[a4paper,11pt]{amsart} 
\usepackage[T1]{fontenc}
\usepackage[english]{babel}
\usepackage{amssymb, amsmath, latexsym}
\usepackage{amsfonts,mathtext,cite,enumerate,float,amsthm}
\usepackage{hyperref}
   \hypersetup{colorlinks,
     linkcolor=blue,
     citecolor=blue,
    urlcolor=blue}
\usepackage{graphicx}
\usepackage{bbm}
\usepackage{mathptmx}
\usepackage{tabularx}
\usepackage{pstricks}
\usepackage{color}

\numberwithin{equation}{section}
\renewcommand{\le}{\leqslant}
\renewcommand{\ge}{\geqslant}

\theoremstyle{definition}

\newtheorem{remark}{\indent Remark}

\usepackage{geometry} 
\author{Petr Chunaev}
\address{National Center for Cognitive Technologies, ITMO University (Saint Petersburg, Russia)}

\title{Community detection in node-attributed social networks: a~survey}

\begin{document}

	\begin{abstract}
		Community detection is a fundamental problem in social network analysis consisting, roughly speaking, in unsupervised dividing social actors (modelled as nodes in a social graph) with certain social connections (modelled as edges in the social graph) into densely knitted and highly related groups with each group well separated from the others. Classical approaches for community detection usually deal only with the structure of the network and ignore features of the nodes (traditionally called node attributes), although the majority of real-world social networks provide additional actors' information such as age, gender, interests, etc. It is believed that the attributes may clarify and enrich the knowledge about the actors and give sense to the detected communities. This belief has motivated the progress in developing community detection methods that use both the structure and the attributes of the network (modelled already via a node-attributed  graph) to yield more informative and qualitative community detection results.
		
		During the last decade many such methods based on different ideas and techniques have appeared. Although there exist partial overviews of them, a recent survey is a necessity as the growing number of the methods may cause repetitions in methodology and uncertainty in practice. 
		
		In this paper we aim at describing and clarifying the overall situation in the field of community detection in node-attributed social networks. Namely, we perform an exhaustive search of known methods and propose a classification of them based on when and how the structure and the attributes are fused. We not only give a description of each class but also provide general technical ideas behind each method in the class. Furthermore, we pay attention to available information which methods outperform others and which datasets and quality measures are used for their performance evaluation. Basing on the information collected, we make conclusions on the current state of the field and disclose several problems that seem important to be resolved in future.
	\end{abstract}
	
\maketitle
	
\tableofcontents
	
\section{Introduction}
	
Community detection is a fundamental problem in social network analysis consisting, roughly speaking, in unsupervised dividing social actors into densely knitted and highly related groups with each group well separated from the others. One class of classical community detection methods mainly deal only with the {\it structure} of social networks (i.e. with connections between social actors) and ignore actors' features. There exist a variety of such structure-aware methods that have shown their efficiency in multiple applications (see \cite{Fortunato2010,Leskovec2010,Lancichinetti2009}). However, the majority of real-world social networks provide more information about social actors than just connections between them. Indeed, it is rather common that certain actors' {\it attributes} such as age, gender, interests, etc., are available. When it is so, the social network is called {\it node-attributed} (recall that the actors are represented via nodes). According to \cite{Wasserman1994}, attributes form the second dimension, besides the structural one, in social network representation. There is another class of classical community detection methods (being opposite to the structure-aware ones, in a sense) that use only node attributes to detect communities and completely ignore connections between social actors. A representative of the attributes-aware methods is well-known $k$-means clustering algorithm taking attribute vectors as an input. Clearly, methods that deal only with structure or only with attributes do not use all the information available in a node-attributed social network. Naturally, this issue can be overcome if a method would somehow jointly use structure and attributes while detecting communities. Developing of such methods became a novel field  in social network analysis \cite{Bothorel2015}. The field is moreover promising as the joint usage is believed to clarify and enrich the knowledge about social actors and to describe the powers that form their communities \cite{Bothorel2015}. 
	
During the last decade numerous methods based on different ideas and techniques have appeared in the field. Although there exist some partial overviews of them, especially in Related Works sections of published papers and in the survey \cite{Bothorel2015} published in 2015, a recent summary of the subject is a necessity as the growing number of the methods may cause repetitions in methodology and uncertainty in practice. 
	
In this survey, we aim at describing and clarifying the overall situation in the field. Namely, we perform an exhaustive search of existing community detection methods for node-attributed social networks. What is more, we propose a classification of them based on when and how they use and fuse network structure and attributes. We not only give a description of each class but also provide general technical ideas behind each method in the class. Furthermore, we pay attention to available information which methods outperform others and which datasets and quality measures are used for their performance evaluation. Basing on the information collected, we make conclusions on the current state of the field and disclose several problems that seem important to be resolved in future. 

To be more precise, let us describe the content of the survey.  In Section~\ref{section2}, we first provide the reader with the notation used in the survey and state the problem of community detection in node-attributed social networks. We further briefly discuss the traditional argumentation in support of such a community detection and the effect of fusing network structure and attributes. In
Section~\ref{section3}, we give information about the related survey works and explain how the search of relevant literature was organized in our case. We also indicate which methods are included in the survey and which are not. Additionally, we explain why references throughout the survey are made in a certain way. Section~\ref{section4} introduces the classification that we propose for community detection methods under consideration. In Section~\ref{section5}, we discuss the most popular datasets and quality measures for evaluation of community detection results. This section is also helpful for simplifying exposition in the forthcoming sections. Sections~\ref{section6}--\ref{section8} contain descriptions of the classes of methods and their representatives. In Section~\ref{section9}, we analyze the overall situation in the field basing on the information from Sections~\ref{section6}--\ref{section8}. Among other things, we disclose several methodological problems that are important to resolve in future studies, in our opinion. Our conclusions on the topic are summarized in Section~\ref{section10}.

\section{Community detection problem for node-attributed social networks and the~effect of fusing network structure and attributes}
\label{section2}
	
\subsection{Necessary notation and the community detection problem statement}
\begin{figure}
	\centering
	\includegraphics[width=0.9\linewidth]{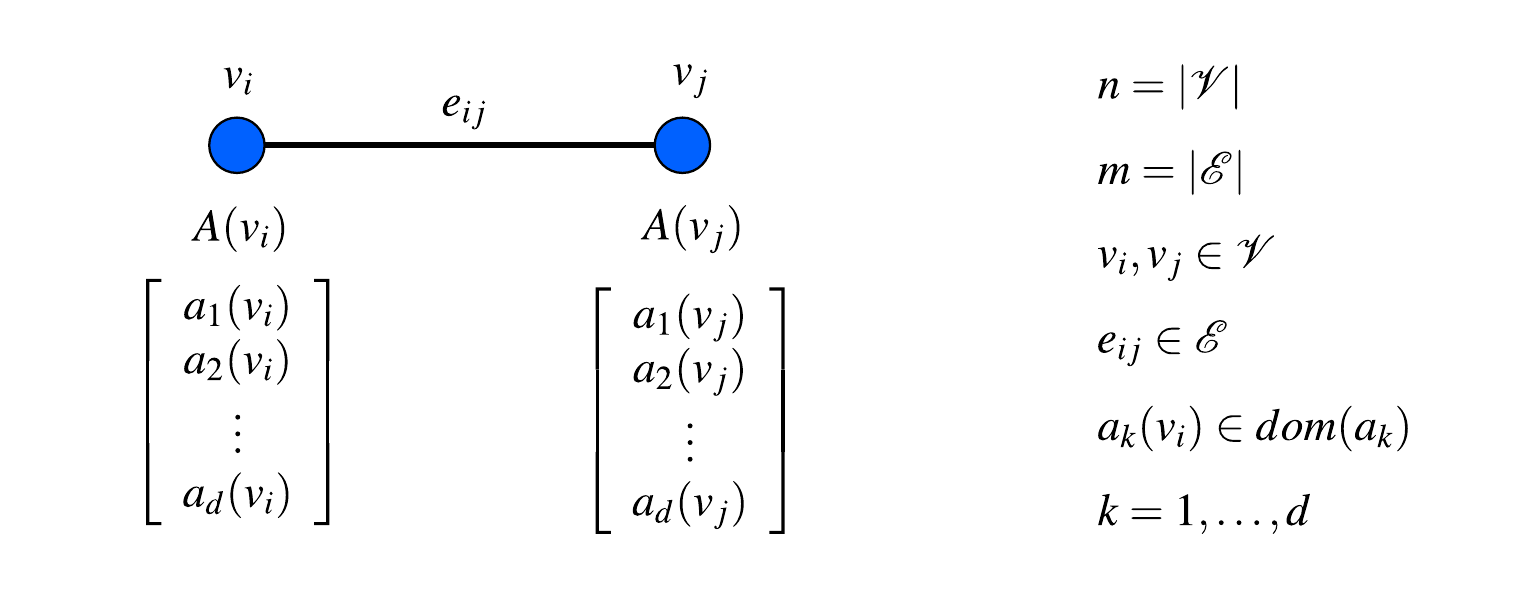}
	\caption{Notation related to the triple (node-attributed graph) $G=(\mathcal{V},\mathcal{E},\mathcal{A})$, where $\mathcal{V}=\{v_i\}$, $\mathcal{E}=\{e_{ij}\}$ and $\mathcal{A}=\{A(v_i)\}$.}
	\label{fig:notation}
\end{figure}

We represent a node-attributed social network as  triple ({\it node-attributed graph}) $G=(\mathcal{V},\mathcal{E},\mathcal{A})$, where $\mathcal{V}=\{v_i\}$ is the set of nodes (vertices) representing social actors, $\mathcal{E}= \{e_{ij}\}$ the set of edges representing connections between the actors ($e_{ij}$ stands for the edge between nodes $v_i$ and $v_j$),  and $\mathcal{A}$ the  set of attribute vectors $A(v_i)=\{a_{k}(v_i)\}$ associated with nodes in  $\mathcal{V}$ and containing information about actors' features. Furthermore, $|\mathcal{V}|=n$, $|\mathcal{E}|=m$ and the dimension of the attribute vectors is $d$. The domain of $a_k$, i.e. the set of possible values of $k$th element of attribute vectors $a_k(v_i)$, is denoted by $dom(a_k)$. In these terms, $k$th attribute of node $v_i$ is referred to as $a_k(v_i)$. The notation introduced above is summarized in Figure~\ref{fig:notation}. Note that pairs $(\mathcal{V},\mathcal{E})$ and $(\mathcal{V},\mathcal{A})$ are correspondingly called the {\it structure} (or {\it topology}) and the {\it attributes} (or {\it semantics}) of  node-attributed graph $G$.

By {\it community detection\footnote{It is also called ``community discovery'' or ``clusterization''.} in node-attributed graph} $G=(\mathcal{V},\mathcal{E},\mathcal{A})$ we mean {\it unsupervised} partitioning the set of nodes $\mathcal{V}$ into $N$ subsets ({\it communities} or {\it clusters}) $C_k\subset \mathcal{V}$, with $C=\{C_k\}_{k=1}^N$ called a {\it partition}, such that
$\mathcal{V}=\bigcup_{k=1}^N C_k$ and a certain balance between the following two properties is achieved:
\begin{itemize}
	\item[(a)] {\it structural closeness}, i.e. nodes within a community are structurally close to each other, while nodes in different communities are not;
	\item[(b)] {\it attribute homogeneity}, i.e.  nodes within a community have  homogeneous attributes, while nodes in different communities do not.
\end{itemize}

Since one can meet variations of the above-mentioned definitions in the relevant literature, it is worth giving several comments on them. First, the number of communities $N$ can be either known in advance or determined during the community detection process automatically. Second, the communities $C_k$ may be defined to be disjoint or overlapping. Third, the property $\mathcal{V}=\bigcup_{k=1}^N C_k$ is sometimes omitted if the resulting partition is not required to include all the nodes from $\mathcal{V}$. Fourth, 
the notion of structural closeness and attribute homogeneity may seem vague at the moment but hopefully become more evident after Subsection~\ref{subsection-str-closeness-attr-homo} and Section~\ref{section4} where reasons and particular measures for them are discussed. Fifth, the definitions given are for the case of nodes and edges each of one type. It is of course possible that social actors and connections between them can be of different types in a social network and thus one should take this heterogeneity into account. This situation is however closer to the notion of multi-layer networks that are out of scope of the present survey (see also Section~\ref{section3}).
	
\subsection{Structural closeness and attribute homogeneity. The effect of fusing  structure and attributes}
\label{subsection-str-closeness-attr-homo}

The structural closeness requirement is based on the recent concepts of a (structural) community in a social network. For example, communities are thought in \cite{Girvan2002} as subsets of nodes with dense connections within the subsets and sparse in between. In its turn, \cite{NewmanGirvan2004} adopts the intuition that nodes within the same community should be better connected than they would be by chance. A possible measure for that is famous Newman's Modularity \cite{Newman2006} that has become an influential tool for structure-based community detection in social networks \cite{Bothorel2015,Chakraborty:2017:MCA:3135069.3091106}. Multiple Modularity modifications and other measures have been also proposed to assess structural closeness \cite{Chakraborty:2017:MCA:3135069.3091106}. In fact, the precise meaning of structural closeness in each community detection method is  determined by the measure chosen.

The attribute homogeneity requirement is based on the social science founding (see e.g. \cite{Marsden1993,McPherson2001,FioreDonath2005,KossinetsWatts2009}) that actors' features  can reflect and affect community structure in social networks. The well-known principle of homophily in social networks states that like-minded social actors have a higher likelihood to be connected \cite{McPherson2001}. Thus community detection process taking into account  attribute homogeneity may provide results of better quality \cite{Bothorel2015}. Oppositely to the situation with structural closeness measures, the attribute homogeneity is usually measured by Entropy that quantifies the degree of disorder of attribute vectors in $(\mathcal{V},\mathcal{A})$ within the communities detected.

Let us now discuss different points of view on the effect of fusing structure and attributes.
From one side, multiple experiments, e.g. in \cite{Moser2009,Ye2017,Sheikh2019,Cohn2001,Getoor2003} and many other papers cited in this survey, suggest that the structure and the attributes of a node-attributed social network often provide complementary information that improves community detection quality. For example, attributes may compensate the structural sparseness of a real-world social network \cite{Jia2017,Yang2013}, while structural information may be helpful in resolving the problem of missing or noisy attributes \cite{Jia2017,Sheikh2019,Yang2013}. What is more, it is observed in \cite{Ding2011} that structure-only or attributes-only community detection is often not as effective as when both sources of information are used. From the other side, some experiments (see e.g. \cite{Akbas2017,Zhou2009}) suggest that this is not always true, and network structure and attributes may be orthogonal and contradictory thus leading to ambiguous community detection results. Moreover, relations between these sources of information may be highly non-linear and challenging to analyze \cite{Wang2016,Yang2013}. 

Besides the above-mentioned points, our general impression is that there is no widely accepted opinion on the effect of fusing structure and attributes and how this fusion can influence community detection quality. Let us illustrate this with an example.
Suppose that the structure of a certain node-attributed social network is ideally correlated with a half of attributes and is wholly orthogonal to another half. For simplicity, let the dimension of attribute vectors be small so that there is no sense to fight against the curse of dimensionality. Now we follow the popular suggestion that the mismatch between structure and  attributes negatively affects community detection quality \cite{Meng2018} and that the existence of structure-attributes correlation offers ``a unique opportunity to improve the learning performance of various graph mining tasks'' \cite{Li2019KDD}. The choice is clear then: we need to use the structure and only the ideally correlated attributes for community detection. It turns out however that  we are going to use two sources of information that mostly duplicate each other. Why should we expect that this improves the quality of detected communities? From our side, we would presume that the structure and the chosen half of attributes (considered separately or jointly) would yield very similar or even the same communities, with all the ensuing consequences for assessing their quality. Shouldn't we use just one of the sources then? Furthermore, it is not clear to us why the other half of attributes should be omitted. Generally speaking, they may contain valuable information for community detection and thus omitting them because of the lack of correlation with the structure is rather questionable.

In any case, a focused theoretical study of when the fusion of structure and attributes is worthy and when not for community detection (ideally, in terms of subclasses of node-attributed social networks) seems to be an extremely important thing that would allow to remove the above-mentioned contradictions.

\section{Related works and processing the relevant literature}
\label{section3}

\subsection{Related works} There is a variety of surveys and comparative studies considering community detection in social networks without attributes, in particular, \cite{SCHAEFFER2007,Yang2016-Survey,Fortunato2010,Coscia2011}. At the same time, the survey 	\cite{Bothorel2015} seems to be the only one on community detection in node-attributed social networks. Obviously, since it was published in 2015, many new methods adapting different fusion techniques have appeared in the field. Furthermore, a big amount of the methods that had been proposed before 2015 are not covered by \cite{Bothorel2015}, in particular, some based on objective function modification, non-negative matrix factorization, probabilistic models, clustering ensembles, etc. In a sense, the technique-based classification of methods in \cite{Bothorel2015} is also sometimes confusing. For example, CODICIL \cite{Ruan2013}, a method based on assigning attribute-aware weights on graph edges, is not included in \cite[Section 3.2. Weight modification according to node attributes]{Bothorel2015}, but to \cite[Section 3.7. Other methods]{Bothorel2015}. Although  \cite{Bothorel2015} is a well-written and highly cited paper, a recent  survey of community detection methods for node-attributed social networks is clearly required.

Besides \cite{Bothorel2015}, almost every paper on the topic contains a Related Works section. It typically has a short survey of preceding methods and an attempt to classify them. We observed that many authors are just partly aware of the relevant literature and this sometimes leads to repetitions in approaches. Furthermore, multiple classifications (usually technique-based) are mainly not full and even contradictory.

\subsection{Relevant literature search process}
\label{subsection 3.2}
At the beginning, we started the search of relevant literature using regular and scientific search engines making the queries like  ``community detection'' or ``clustering'' or
``community discovery'' in ``node-attributed social networks'' or ``node-attributed graphs''. Within the search process it became evident that other queries also lead to the relevant literature. In particular, ``clustering an attribute-aware graph'', ``community detection in networks with node attributes'', ``description-oriented community detection'', ``semantic clustering of social networks'', ``structure and attributes community detection'', ``joint cluster analysis of attribute and relationship data'', ``community discovery integrating links and tags'', ``attributed graph partitioning'', ``node attribute-enhanced community detection'', ``community detection based on structure and content'', etc. It can be also observed that node-attributed networks and graphs are also sometimes called ``augmented networks'', ``graphs with feature vectors'', ``feature-rich networks'' and ``multi-attributed graphs''.
This variety of terms suggests that there is still no established terminology in the field and emphasizes the significance of our survey, where we try to use consistent terminology.

After the above-mentioned exhaustive search, we learned the references in the found papers.  Among other things, it brought us to ideologically close papers devoted, for example, to ``attributed information networks'', ``annotated document networks'', ``multi-layer networks'' and ``subspace-based clustering''. We stopped further search when we could not find any new relevant references. Since this happened in the middle of 2019, the survey covers the found papers that had been published in journals or conference proceedings before this date. 

\subsection{The format of references to methods and datasets} 
It turns out that several methods for community detection in node-attributed social networks can be proposed in one paper. Therefore, a regular reference of the form [ReferenceNumber] may be not informative enough. From the other hand, authors usually provide their methods with short names like SA-Cluster, CODICIL or CESNA\footnote{If this is not the case, we allowed ourselves to invent our own names suggesting the class the method belongs to in our classification.}. Some of the names are rather familiar to researches in the field. Thus it seems reasonable to make a reference of the form MethodName [ReferenceNumber] and so we do in what follows. However, not all the methods that are mentioned in the survey are included in our classification and thus discussed in a more detailed manner (as some are just out of scope). To distinguish the cases, we write names of the methods included in our classification in {\bf bold}, e.g. {\bf SCMAG} \cite{Huang2015}, {\bf UNCut} \cite{Ye2017} and {\bf DCM} \cite{Pool2014}. Such a format means that the reader can find short descriptions of the methods {\bf SCMAG} \cite{Huang2015}, {\bf UNCut} \cite{Ye2017} and {\bf DCM} \cite{Pool2014} in our survey. References like DB-CSC \cite{Gunnemann2011}, FocusCO \cite{Perozzi2014-2} and ACM \cite{Wu2018} mean that the corresponding methods are not included in our classification. In this case the reader is recommended to go directly to the papers \cite{Gunnemann2011}, \cite{Perozzi2014-2} and \cite{Wu2018} to get additional information.  

A similar scheme is applied to names of the node-attributed network datasets discussed in Section~\ref{section5} and used in further classification. The reader is assumed to have in mind that if a dataset name is written in {\bf bold}, then its description can be found in Tables~\ref{tab_dataset1}, \ref{tab_dataset2} or \ref{tab_dataset3}. Note also that various versions of the datasets from Tables~\ref{tab_dataset1}, \ref{tab_dataset2} or \ref{tab_dataset3} are in fact used in different papers. To show that a dataset is somehow different from the description in Tables~\ref{tab_dataset1}, \ref{tab_dataset2} or \ref{tab_dataset3}, we mark it by *. For example, a DBLP dataset with the number of nodes and edges different from {\bf DBLP10K} and {\bf DBLP84K} in Table~\ref{tab_dataset2} is denoted by {\bf DBLP*} in Sections~\ref{section6}--\ref{section8}.

\subsection{Note on multi-layer network clustering} In the survey, we do not  consider community detection methods for multi-layer networks, where different types of vertices and edges may present at different layers \cite{Kivela2014,Interdonato2019}. Nevertheless, we mention some of these methods from time to time in corresponding remarks. It is though important to note that node-attributed networks of different nature may be clearly considered as a particular case of multi-layer ones. In the majority of papers covered by the present survey, this connection is however rarely commented. 

Let us also emphasize that  multi-layer networks (graphs) require special analysis taking into account the heterogeneity of   vertices and edges on different layers. A~separate survey and an extensive comparable study of such methods is an independent and useful task (see partial overviews e.g. in \cite{Boutemine2017,Interdonato2019,Kivela2014}).

\subsection{Note on subspace-based clustering}
Following the above-mentioned definition of community detection in node-attributed social networks, we mainly consider in the survey the methods that can use the full attribute space and find communities covering the whole network. However, there is a big class of special methods that explore subspaces of attributes and/or deal with  subgraphs of an initial graph, e.g. GAMer \cite{Gunnemann2010,Gunnemann2014}, DB-CSC \cite{Gunnemann2011}, SSCG \cite{Gunnemann2013}, FocusCO \cite{Perozzi2014-2} and ACM \cite{Wu2018}.
The main idea behind the subspace-based (also known as projection-based) clustering methods is that not all available semantic information (attributes) is relevant to obtain good-quality communities \cite{Gunnemann2013-1,GunnemannBoden2013}. For this reason, one has to somehow choose the appropriate attribute subspace to avoid the so-called {\it curse of dimensionality} (see \cite[Section 3.2]{Bothorel2015}) and reveal significant communities that would not be detected if all available attributes were considered.

To be precise, some of the methods that we discuss below partly use this idea, e.g. {\bf WCru} \cite{Cruz2011,CruzBathorelPoulet2012} (cf. the definition of a {\it point of view} in the papers), {\bf DVil} \cite{Villavialaneix2013}, {\bf SCMAG} \cite{Huang2015}, {\bf UNCut} \cite{Ye2017}, {\bf DCM} \cite{Pool2014}, etc., but still can work with the full attribute space. In any case, a separate survey on  subspace-based  methods for community detection in node-attributed social networks would be a very  valuable complement to the current one.

\subsection{Note on community detection in node-attributed networks of different type and its applications} 
Clearly, community detection tools for node-attributed social networks are suitable for networks of different nature. That is why, besides obvious  applications in marketing (recommender systems, targeted advertisements and user profiling) \cite{Alamsyah2014},  the tools are used for search engine
optimization and spam detection \cite{Ruan2013,Muslim2016},  in counter-terrorist activities and disclosing fraudulent schemes \cite{Muslim2016}. They are also applied to  analysis of protein-protein interactions, genes and epidemics \cite{Muslim2016}.

Another possible application is document network clustering. Note that such a clustering is historically preceding to  community detection in node-attributed social networks and is rich methodologically on its own \cite{Nail2016,Aggarwal2012,Saiyad2016}. One should take into account however that social communities although have similar formal description with document clusters   have inner and more complicated forces to be formed and act. What is more, it has been shown that methods for community detection in node-attributed social networks outperform preceding methods for document network clustering in many cases, see \cite{Zhou2010,Yang2009,Yang2013,Balasubramanyan2011}. For this reason, we do not consider document network clustering methods in this survey.
	
\section{Our classification of community detection methods for~node-attributed~ social~networks}
\label{section4}

\begin{figure}[b]
\begin{minipage}[t]{0.325\textwidth}
    \includegraphics[width=\textwidth]{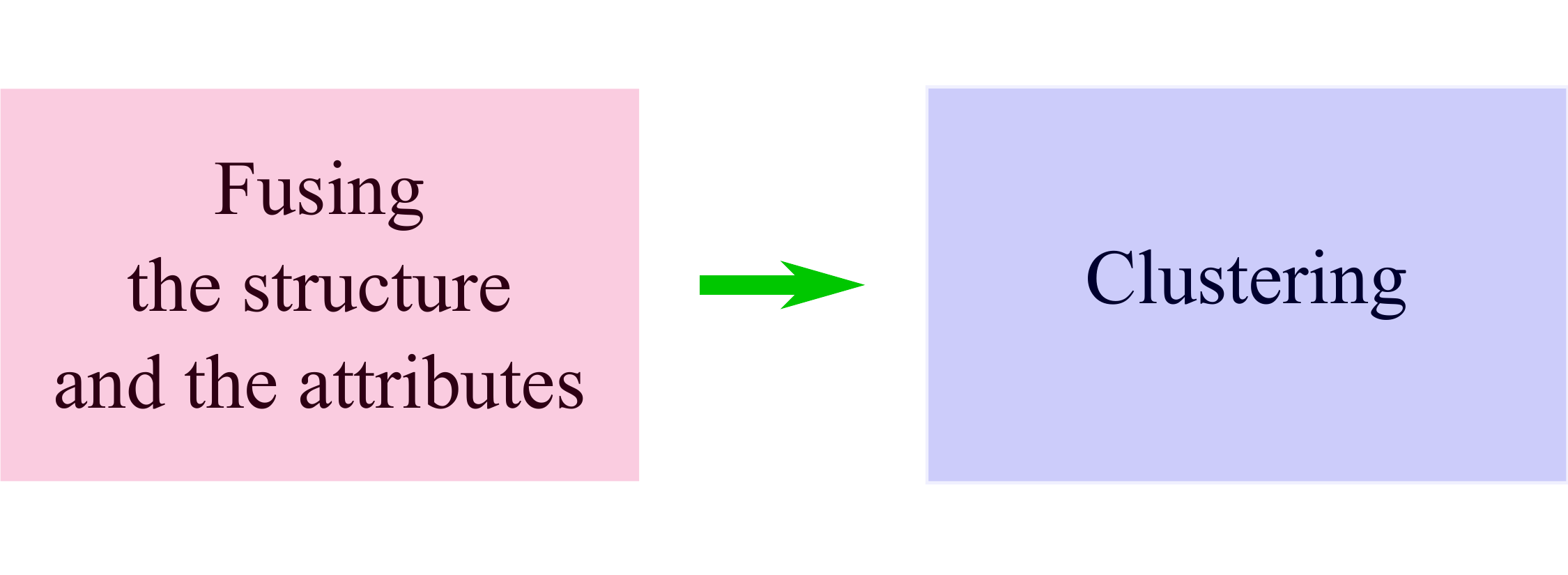}
    
    \centering 
    
    {\bf Early fusion methods}\\ (Section~\ref{section6})
    
    \vspace{0.3cm}
    
    \hrule
    
    \vspace{0.3cm}
    
         Weight-based (Section~\ref{class-ef-weight}) 
         
         \vspace{0.3cm}
         
         Distance-based (Section~\ref{class-ef-distance})
         
         \vspace{0.3cm}
         
         Node-augmented graph-based (Section~\ref{class-ef-node-augmented-graph}) 
         
         \vspace{0.3cm}
         
         Embedding-based (Section~\ref{class-ef-embedding})
         
         \vspace{0.3cm}
         
         Pattern mining-based (Section~\ref{class-ef-patterns})
         
\end{minipage}
\begin{minipage}[t]{0.325\textwidth}
    \includegraphics[width=\textwidth]{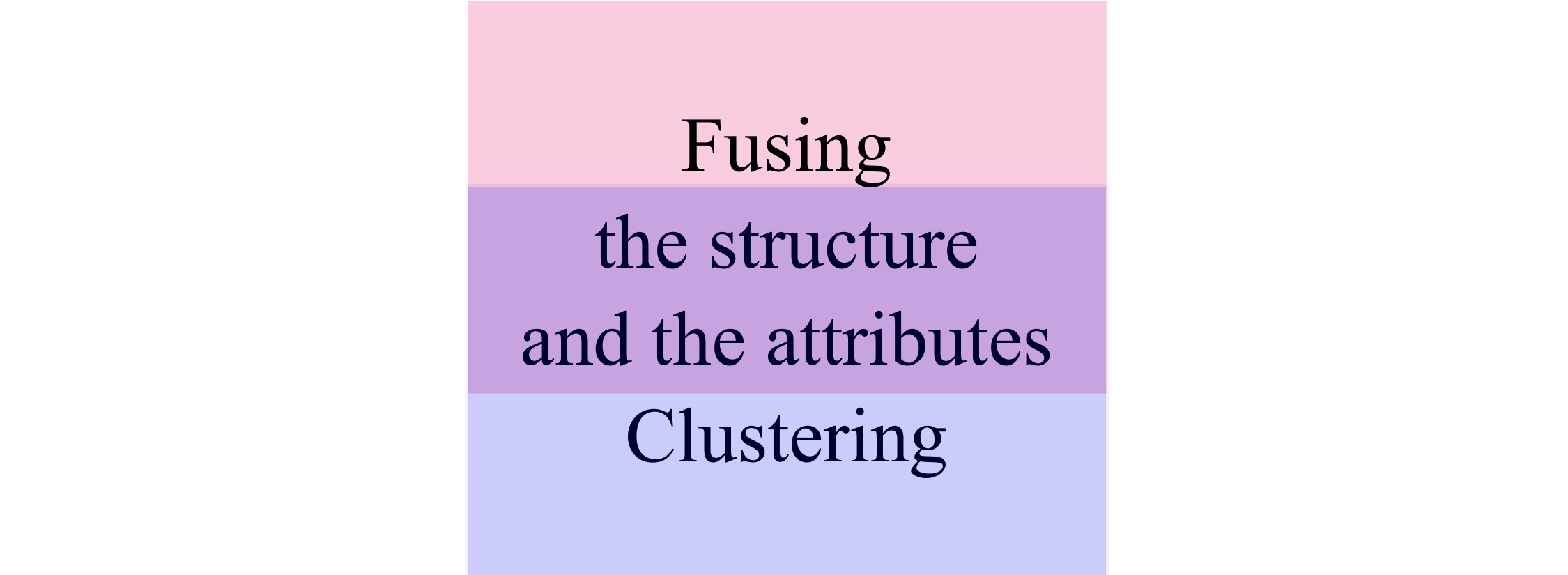}
    \centering 
    
    {\bf Simultaneous fusion methods} (Section~\ref{section7})
    
    \vspace{0.3cm}
    
    \hrule
    
    \vspace{0.3cm}
    
     Methods modifying objective functions of classical clustering algorithms (Section~\ref{class-sf-modifying}) 
     
     \vspace{0.3cm}
     
     Metaheuristic-based (Section~\ref{class-sf-mataheuristic})
     
     \vspace{0.3cm}
     
     Methods based on non-negative matrix factorization and matrix compression (Section~\ref{class-sf-nnmf}) 
     
     \vspace{0.3cm}
     
     Pattern mining-based (Section~\ref{class-sf-patterns}) 
     
     \vspace{0.3cm}
     
     Probabilistic model-based (Section~\ref{class-sf-probabilistic-models})  
     
     \vspace{0.3cm}
     
     Dynamical system-based and agent-based (Section~\ref{class-sf-dynamics})

\end{minipage}
\begin{minipage}[t]{0.325\textwidth}
    \includegraphics[width=\textwidth]{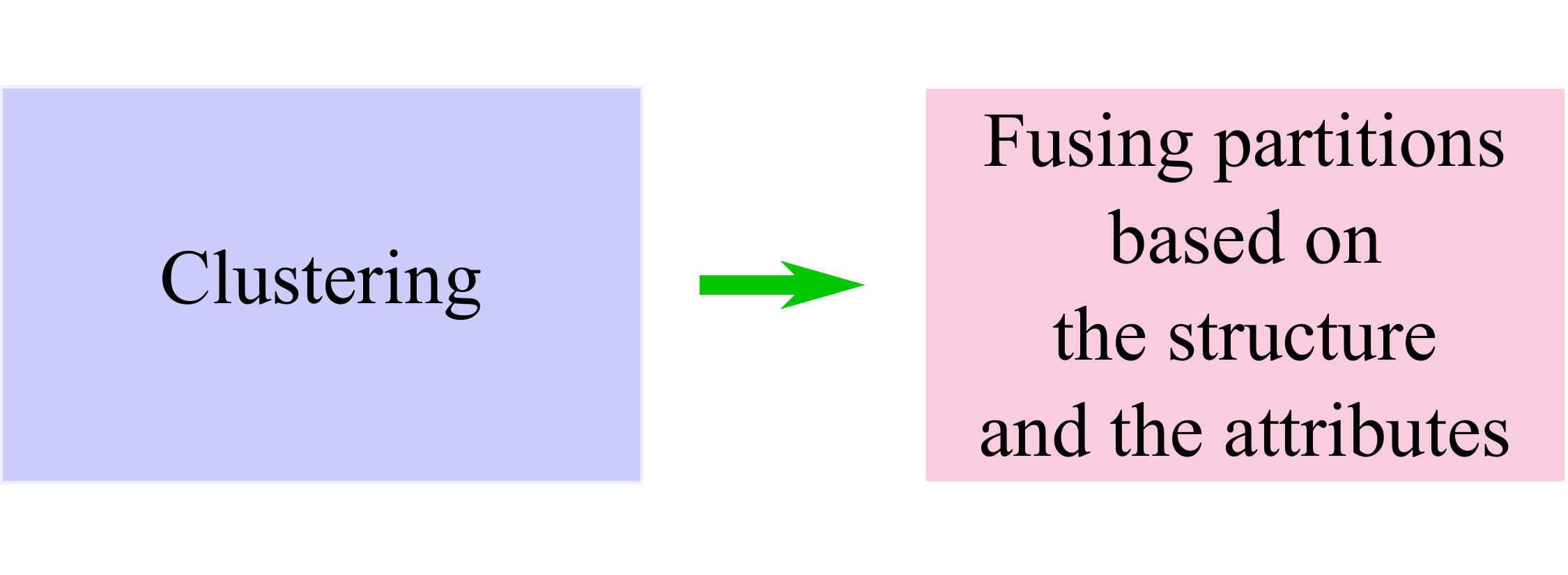}
    \centering
    
    {\bf Late fusion methods}
    
    (Section~\ref{section8})
    
    \vspace{0.3cm}
    
    \hrule
    
    \vspace{0.3cm}
    
     Consensus-based (Section~\ref{class-lf-consensus})
     
     \vspace{0.3cm}
     
     Switch-based (Section~\ref{class-lf-switch})
     
\end{minipage}
    \caption{The proposed classification and the survey structure guide.}
    \label{fig:pics_clafficication}
\end{figure}

In previous works, the classification of methods for community detection in node-attributed social networks was done mostly with respect to the techniques used (e.g. distance-based or random walk-based). We partly follow this principle but at a lower level. At the upper level we group the methods by when  structure and attributes are fused with respect to the community detection process, see Figure~\ref{fig:pics_clafficication}. Namely, we distinguish the classes of 
	\begin{itemize}
		\item {\it early fusion methods} that fuse structure and attributes before the community detection process,
		\item {\it simultaneous fusion methods} that fuse structure and attributes simultaneously with the community detection process,
		\item {\it late fusion methods} that first partition structure and attributes  separately and further fuse the partitions obtained.
	\end{itemize}
	
Such a classification allows an interested researcher/data scientist to estimate the labour costs of software implementation of the method chosen for use in practice.
Indeed, early fusion methods require just a preprocessing (fusion) procedure converting information about  structure and attributes into a format that is often suitable for classical community detection algorithms. For example, weight-based early fusion methods convert attribute vectors into the graph form and further merge the structure- and attributes-aware graphs into a weighted graph that can be processed by graph clustering algorithms with existing implementations. Implementation of late fusion methods is also rather simple. Namely, at the first step they detects communities by classical graph and vector clustering algorithms applied to structure and attributes separately. At the second step, the communities obtained are somehow merged by an existing (or at least easy-implemented) algorithm. Oppositely to the early and late fusion methods, simultaneous fusion ones require more programmer work as usually assume either a completely new implementation or essential modifications to existing ones. The situation is aggravated by the fact that their source codes are rarely available.

As seen from Figure~\ref{fig:pics_clafficication}, we also divide the methods within each class into technique-used subclasses. Let us emphasize that the priority in this division is the fusion technique.  For example, by ``weight-based methods'' we mean those which form a weighted graph while fusing structure and attributes. The majority of such methods further use graph clustering algorithms for community detection (and this is reasonable). However, some may still transform the graph into the distance matrix form and then use distance-based clustering algorithms. Such methods are  still called ``weight-based''. One more example: ``distance-based methods'' are called in this way as directly produce a distance matrix while fusing structure and attributes, independently on how this matrix is further processed.

What is more, we provide short descriptions of methods in each class and subclass in tables in Sections~\ref{section6}--\ref{section8}. In particular, we briefly describe the community detection algorithm and its input used in the method,  and the type of communities obtained (overlapping or not). Furthermore, we mention which datasets and quality measured are used by method's authors for evaluation of community detection results. In addition, for each method  we provide a list of other methods for {\it community detection in node-attributed networks} (using both structure and attributes) that the method under consideration is compared with. Note that the list may be empty sometimes. This is so, for example, if the method under consideration is  compared only with classical community detection methods that deal either with structure or  attributes and do not fuse them to detect communities.

\section{Most used node-attributed network datasets and  quality measures for community detection evaluation}
\label{section5}

\subsection{Datasets}

The title of the survey suggests that we are focused on community detection in node-attributed {\it social} networks. However, the methods that are included in our classification, generally speaking, may be applied to node-attributed networks of different nature. As we have noticed, authors of the methods implicitly share this point of view and freely use various node-attributed network datasets to evaluate community detection quality.

In this subsection (Tables~\ref{tab_dataset1}, \ref{tab_dataset2} or \ref{tab_dataset3}) we collect and briefly describe the datasets that are popular in the field\footnote{An interested reader can find more node-attributed network datasets at \href{http://www-personal.umich.edu/~mejn/netdata/}{Mark Newman page},  \href{https://hpi.de/naumann/projects/repeatability/datasets/}{HPI Information Systems Group}, \href{https://linqs.soe.ucsc.edu/data}{LINQS Statistical Relational Learning Group}, \href{http://snap.stanford.edu/data/}{Stanford Large Network Dataset Collection}, \href{http://dp.univr.it/~laudanna/LCTST/downloads/index.html}{University of Verona Laboratory of Cell Trafficking and Signal Transduction}, \href{https://perso.liris.cnrs.fr/marc.plantevit/doku/doku.php?id=data_sets}{Marc Plantevit page}, \href{https://toreopsahl.com/datasets/}{Tore Opsahl page}, \href{https://sites.google.com/site/ucinetsoftware/datasets}{UCINET networks}, \href{http://networkrepository.com}{Interactive Scientific Network Data Repository}, \href{https://aminer.org/citation}{Citation Network Dataset}.
}. Recall that the dataset names written in {\bf bold} in Sections~\ref{section6}--\ref{section8} refer to the tables in this subsection.  
It can be observed that Tables~\ref{tab_dataset1}, \ref{tab_dataset2} or \ref{tab_dataset3} contain datasets based on social network data (e.g. Facebook, LastFM and Twitter) and document or citation network data (e.g. DBLP, Wiki and Patents). 

For convenience, we distinguish the datasets by size. Namely, by {\it small}, {\it medium} and {\it large} we mean network datasets with $< 10^3$, $10^3\ldots 10^5$ and $>10^5$ nodes, correspondingly.

\begin{table}
	\caption{Most popular small size datasets.}
	\label{tab_dataset1} 
	{\tiny
		\begin{tabular}{|p{1.2cm}|p{12.9cm}|p{0.8cm}|} 
			\hline 
			Dataset  & Description & Source\\ 
			\hline 
			{\bf Political Books}   & All books in this dataset were
			about U.S. politics published during the 2004 presidential election
			and sold by Amazon.com. Edges between books means
			two books are always bought together by customers. Each
			book has only one attribute termed as political persuasion, with
			three values: 1) conservative; 2) liberal; and 3) neutrality & \href{http://www-personal.umich.edu/~mejn/netdata/}{Link} \\ 
			\hline 
			{\bf WebKB} & A classified network of 877 webpages (nodes) and 1608 hyperlinks (edges) gathered from four different universities Web sites (Cornell, Texas, Washington, and Wisconsin). Each web page is associated with a binary vector, whose elements take the value $1$ if the corresponding word from the vocabulary is present in that webpage, and $0$ otherwise. The vocabulary consists of 1703 unique words. Nodes are classified into five classes: course, faculty, student, project, or staff. & \href{https://linqs.soe.ucsc.edu/data}{Link} \newline \cite{Craven1998} \\ 
			\hline 
			{\bf Twitter}  & A collection of several tweet networks: 1) Politics-UK dataset is collected from Twitter accounts of 419 Members of Parliament in the United Kingdom in 2012. Each user has 3614-dimensional attributes, including a list of words repeated more than 500 times in their tweets. The accounts are assigned to five
			disjoint communities according to their political affiliation.
			2) Politics-IE dataset is collected from 348 Irish
			politicians and political organizations, each user has 1047-
			dimensional attributes. The users are distributed into seven
			communities. 3) Football dataset contains 248 English
			Premier League football players active on Twitter which are
			assigned to 20 disjoint communities, each corresponding to
			a Premier League club. 4) Olympics dataset contains users
			of 464 athletes and organizations involved in the London
			2012 Summer Olympics. The users are grouped into 28
			disjoint communities, corresponding to different Olympic
			sports. & \href{http://mlg.ucd.ie/aggregation/}{Link 1} \newline \href{http://mlg.ucd.ie/networks/}{Link 2}\newline \cite{Greene2013}    \\ 
			\hline 
		{\bf Lazega}   & A corporate law partnership in a Northeastern US corporate law firm; possible attributes: (1: partner; 2: associate), office (1: Boston; 2: Hartford; 3: Providence); 71 nodes and 575 edges & \cite{Lazeda2001}  \\ 
			\hline 
			{\bf Research}  & A research team of employees in a manufacturing company; possible attributes: location (1: Paris; 2: Frankfurt; 3: Warsaw; 4: Geneva), tenure (1: 1--12 months; 2: 13--36 months; 3: 37--60 months; 4: 61+ months); 77 nodes and 2228 edges & \cite{Cross2004}  \\ 
			\hline 
			{\bf Consult}  & Network showing relationships between employees in a consulting company; possible attributes: organisational level (1: Research Assistant; 2: Junior Consultant; 3: Senior Consultant; 4: Managing Consultant; 5: Partner), gender (1: male; 2: female); 46 nodes and 879 edges & \cite{Cross2004}  \\   
			\hline 
		\end{tabular}
	}
\end{table} 
\begin{table}
	\caption{Most popular medium size datasets.}
	\label{tab_dataset2} 
	{\tiny
		\begin{tabular}{|p{1.2cm}|p{12.9cm}|p{0.8cm}|}
			\hline 
			Dataset  & Description & Source \\ 
			\hline 
			{\bf Political Blogs}  & A non-classified network of 1,490 webblogs (nodes) on US politics with
			19,090 hyperlinks (edges) between the webblogs. Each node has an attribute describing its political leaning as either liberal or
			conservative (represented by $0$ and $1$). & \href{http://www-personal.umich.edu/~mejn/netdata/}{Link}\newline \cite{Adamic2005}  \\ 
			\hline 
			{\bf DBLP10K} &  A non-classified co-author network extracted from DBLP Bibliography (four research areas of database, data mining, information retrieval and artificial intelligence) with 10,000  authors (nodes) and their co-author relationships (edges). Each author is associated with two relevant categorical attributes: prolific and primary topic. For attribute ``prolific'', authors with $\ge 20$ papers are labelled as highly prolific; authors with $>10$ and $<20$ papers are labelled as prolific and authors with $\le 10$ papers are labelled as low prolific. Node-attribute  values for ``primary topic'' (100 research topics) are obtained via topic modelling. Each extracted topic consists of a probability distribution of keywords which are most representative of the topic. & \href{https://github.com/Issamfalih/ANCL/tree/master/data}{Link}\newline \cite{Zhou2010} \\ 
			\hline 
			{\bf DBLP84K} &  A larger non-classified co-author network extracted from DBLP Bibliography (15 research areas of database, data mining, information retrieval, artificial intelligence,  machine learning, computer vision, networking, multimedia, computer systems, simulation, theory, architecture, natural language processing, human-computer interaction, and programming language) with 84,170  authors (nodes) and their co-author relationships (edges). Each author is associated with two relevant categorical attributes: prolific and primary topic, defined in a similar way as in DBLP10. &  \href{https://github.com/Issamfalih/ANCL/tree/master/data}{Link}\newline \cite{Zhou2010} \\ 
			\hline 
			{\bf Cora}  & A classified network of machine learning papers with 2,708 papers (nodes) and 5,429 citations (edges). Each node is attributed with a $1433$-dimension binary vector indicating the absence/presence of words from the dictionary of words collected from the corpus of papers. The papers are  classified into 7 subcategories: case-based reasoning,genetic algorithms,
			neural networks, probabilistic methods, reinforcement learning,
			rule learning and theory.	 &  \href{https://linqs.soe.ucsc.edu/data}{Link 1}\newline \href{https://github.com/albertyang33/TADW}{Link 2}\newline \cite{Sen2008} \\ 
			\hline 
			{\bf CiteSeer}  & A classified citation network in the field of machine learning with 3,312 papers (nodes) and 4,732 citations (edges). Each node is attributed with a binary vector indicating the absence/presence of the corresponding words from the dictionary of the 3,703 words collected from the corpus of papers. Papers are classified into 6 classes. & \href{https://linqs.soe.ucsc.edu/data}{Link 1}\newline \href{https://github.com/albertyang33/TADW}{Link 2}\newline \cite{Sen2008}\\ 
			\hline 
			{\bf Sinanet}  & A classified microblog user relationship network extracted from the sina-microblog website (http://www.weibo.com) with 3,490 users (nodes)  and 30,282 relationships (edges). Each node is attributed with 10-dimensional numerical attributes describing the interests of the user. & \href{https://github.com/smileyan448/Sinanet}{Link}\newline \cite{Jia2017} \\ 
			\hline 
			{\bf PubMed Diabetes}  & A classified citation networks extracted from the PubMed database pertaining to diabetes. It contains  19,717 publications (nodes) and 44,338 citations (edges). Each node is attributed by a TF-IDF weighted word vector from a dictionary that consists of 500 unique words. & \href{https://linqs.soe.ucsc.edu/data}{Link} \\ 
			\hline
			{\bf Facebook100}   & A non-classified Facebook users network with 6,386 users (nodes) and 435,324 friendships (edges). The network is gathered from Facebook users of 100 colleges and universities (e.g. Caltech, Princeton, Georgetown and UNC Chapel Hill) in September 2005. Each user has the following attributes: ID, a student/faculty status flag, gender, major, second major/minor (if applicable), dormitory(house), year and high school. & \href{https://archive.org/download/oxford-2005-facebook-matrix}{Link}\newline \cite{Traud2012,Traud2011}  \\ 
			\hline
			{\bf ego-Facebook}   & Dataset consists of 'circles' ('friends lists') from Facebook with 4039 nodes and 88234 edges. Facebook data was collected from survey participants using a Facebook app. The dataset includes node features (profiles), circles, and ego networks. &  \href{https://snap.stanford.edu/data/ego-Facebook.html}{Link}\newline \cite{Leskovec2012}\\ 
			\hline 
			{\bf LastFM}   & A network gathered from the online music system Last.fm with 1,892 users (nodes) and 12,717 friendships on Last.fm (edges). Each node has 11,946-dimensional attributes, including a list of most listened music artists, and tag assignments.
			& \href{http://ir.ii.uam.es/hetrec2011/datasets.html}{Link} \\ 
			\hline 
			{\bf Delicious}  & A network of 1,861 nodes, 7,664 edges and 1,350 attributes. This is a publicly available dataset from the HetRec 2011 workshop that has been obtained from the Delicious social bookmarking system. Its users are connected in a social network generated from Delicious mutual fan relations. Each user has bookmarks, tag assignments, that is, [user, tag, bookmark] tuples, and contact relations within the social network. The tag assignments were
			transformed to attribute data by taking all tags that a user ever assigned to any bookmark and assigning those to the user. & \href{http://ir.ii.uam.es/hetrec2011/}{Link} \\ 
			\hline 
			{\bf Wiki}  & A network with nodes as web pages. The link among different nodes is the hyperlink in the web page. 2,405 nodes, 12,761 edges, 4,973 attributes, 17 labels & \href{https://github.com/albertyang33/TADW}{Link} \\ 
		\hline 
		{\bf ego-Twitter}  & This dataset consists of 'circles' (or 'lists') from Twitter. Twitter data was crawled from public sources. The dataset includes node features (profiles), circles, and ego networks. Nodes	81306,
		Edges 1768149 & \href{https://snap.stanford.edu/data/ego-Twitter.html}{Link}\newline \cite{Leskovec2012} \\
		\hline 
		\end{tabular}
	}
\end{table} 

\begin{table}
	\caption{Most popular large size datasets.}
	\label{tab_dataset3} 
		{\tiny
\begin{tabular}{|p{1.2cm}|p{12.9cm}|p{0.8cm}|}
	\hline 
Dataset  & Description & Source \\ 
	\hline 
{\bf Flickr}  &  A network with 100,267 nodes, 3,781,947 edges and 16,215 attributes collected from the internal database of the popular Flickr
photo sharing platform. The social network is defined by the contact relation of
Flickr. Two vertices are connected with an undirected edge if at least one undirected edge exists between them. Each user has a list of tags associated that he/she used at least five times. Tags are limited to those used by at least 50 users. Users are limited to those having a vocabulary of more than 100 and less than 5,000 tags. &  \href{https://snap.stanford.edu/data/web-flickr.html}{Link}\newline \cite{Ruan2013}\\ 
\hline 
{\bf Patents}  & A patent citation network with vertices representing patents and edges depicting the citations between. A subgraph containing all the patents from the year 1988 to 1999. Each patent has six attributes, grant year, number of claims, technological category, technological subcategory, assignee type, and main patent class. There are 1,174,908 vertices and 4,967,216 edges in the network.  & \href{http://www.nber.org/patents/}{Link 1}\newline \href{https://snap.stanford.edu/data/cit-Patents.html}{Link 2}  \\ 
\hline
{\bf ego-G+} & This dataset consists of 'circles' from Google+. Google+ data was collected from users who had manually shared their circles using the 'share circle' feature. The dataset includes node features (profiles), circles, and ego networks. Nodes 107,614, Edges 13,673,453. Each node has four features: job
title, current place, university, and workplace. A user-pair(edge) is
compared using knowledge graphs based on, Category: Occupations,
Category:Companies by country and industry, Category: Countries,
Category:Universities and colleges by country. & \href{https://snap.stanford.edu/data/ego-Gplus.html}{Link}\newline \cite{Leskovec2012} \\
\hline  
\end{tabular}
}
\end{table} 

\subsection{Community detection quality measures}
Given a set of detected communities (overlapping or not), one needs to evaluate their quality. There are two possible options for this depending on the  network dataset under consideration. If the network dataset has no ground truth, one can use various measures of structural closeness and attribute homogeneity. According to our observations, the most popular quality measures in this case are Modularity and Density for the former and Entropy for the latter. Many others such as Conductance, Permanence, Intra- and Inter-Cluster Densities, etc., are also possible. If there is ground truth, it is traditional to compare the detected communities with the ground truth ones. This can be done, for instance, with the following popular measures: Accuracy, Normalized Mutual Information (denoted below by NMI), Adjusted Rand Index or Rand Index (denoted below by ARI and RI, correspondingly) and $F$-measure. We will discuss both the approaches further in Section~\ref{section9}.  

Due to space limitations, we refer the reader to the comprehensive survey \cite{Chakraborty:2017:MCA:3135069.3091106} and to \cite[Sections 2.2 and 4]{Bothorel2015}, where all the above-mentioned quality measures and many others are precisely defined and discussed in detail.

\section{Early fusion methods}
\label{section6}
As we have already mentioned, these methods fuse structure and attributes before the community detection process so that the data obtained are suitable for classical community detection algorithms (thus one can use their existing software implementations).

Before we proceed, we introduce additional notation applied only to the weight-based and distance-based early fusion methods. The fact is that existing network structure may be saved or modified depending on the heuristics used in a method, therefore we distinguish
	\begin{itemize}
		\item {\it fixed topology methods} that use initial network structure without modifying it with respect to attributes,
		\item {\it non-fixed topology methods} that modify initial network structure with respect to attributes, in particular, add/erase edges and/or vertices.
	\end{itemize}
	
As far as we know, there is no study on which approach is preferable. How each one influences community detection results is yet to be established.

\begin{figure}[b]
	\centering
	\includegraphics[width=0.9\linewidth]{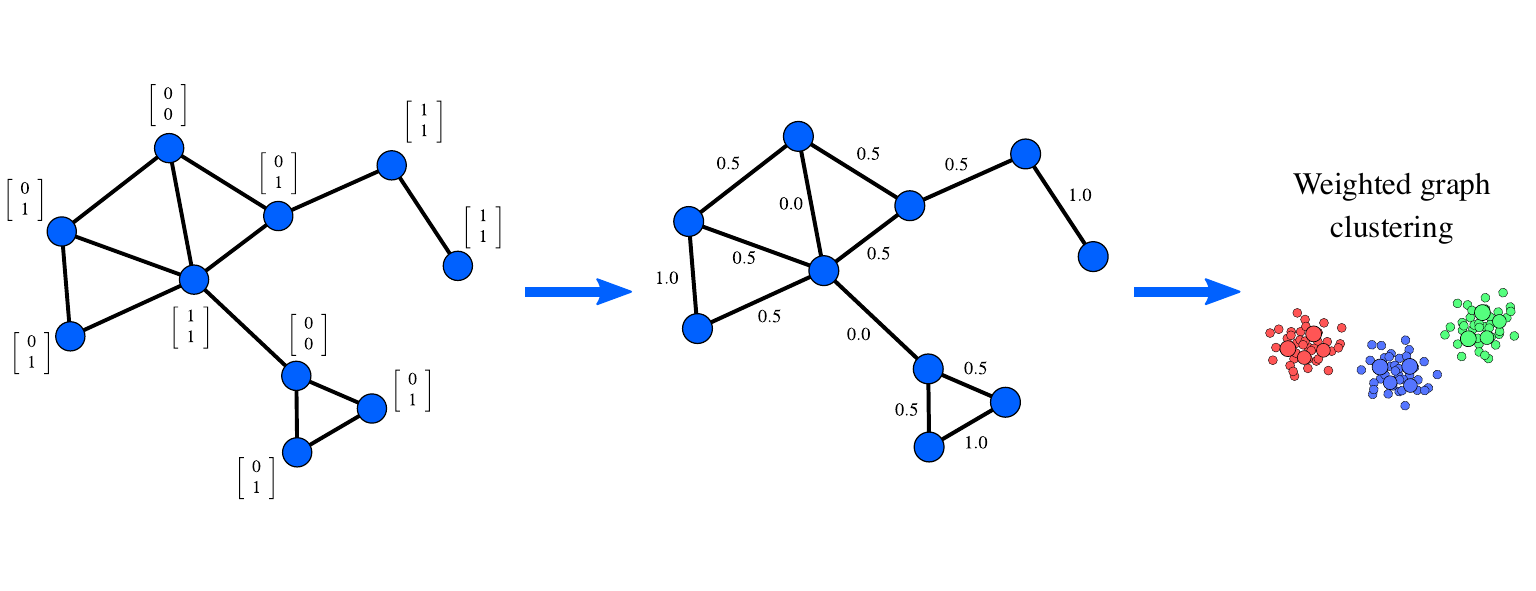}
	\caption{A typical scheme of a weight-based method (the attributive weights here are the values of normalized matching coefficient for the attribute vectors).}
	\label{fig:weight}
\end{figure}

\subsection{Weight-based methods}
\label{class-ef-weight}
These methods (see Tables \ref{tab_weights_alpha0} and \ref{tab_weights}) convert attributes $(\mathcal{V},\mathcal{A})$ into a weighted {\it attributive} graph and further somehow merge it with structural graph $(\mathcal{V},\mathcal{E})$.
The result is a weighted graph $G_W$ that is a substitution for node-attributed graph $G$, see Figure~\ref{fig:weight}. Edge weights of $G_W$ are usually assigned as follows:
\begin{equation}
\label{main_weight}
W_\alpha(v_i,v_j)=\alpha w_S(v_i,v_j)+(1-\alpha) w_A (v_i,v_j),\qquad \alpha\in[0,1],\qquad v_i,v_j\in \mathcal{V},
\end{equation}
where $w_S$ and $w_A$ are chosen {\it structural} and  {\it attributive}  similarity functions, respectively. The hyperparameter $\alpha$ controls the balance between structure and attributes. Clearly,  if $\alpha=1$ in (\ref{main_weight}), one obtains weights based only on  structure; if $\alpha=0$, then they are based only on  attributes. As for $w_S$ and $w_A$, $w_S(v_i,v_j)$ usually reflects existing connections in $(\mathcal{V},\mathcal{E})$ (e.g. $w_S(v_i,v_j)=1$, if $e_{ij}\in \mathcal{E}$, and $w_S(v_i,v_j)=0$, otherwise),  while $w_A(v_i,v_j)$ may be Cosine Similarity or Matching Coefficient values for vectors $A(v_i)$ and $A(v_j)$.

Once the weighted graph $G_W$ is constructed, one can use classical graph clustering algorithms on it such as Weighted  Louvain \cite{Blondel2008}. Sometimes $G_W$ is instead converted to a certain distance matrix that is further used for detecting communities via distance-based clustering algorithms such as  $k$-means or $k$-medoids.

It is worth mentioning that the fixed topology methods in this subclass assume that  the weights (\ref{main_weight}) are assigned on the same set of edges $\mathcal{E}$ as in the initial node-attributed graph $G$, while the non-fixed topology ones assign weights on all (or on the most part of) possible edges between nodes in $\mathcal{V}$.

Now let us say some words how the hyperparameter $\alpha$ can be chosen. A very popular approach in fixed topology methods is assuming $\alpha=0$ in (\ref{main_weight}), see Table~\ref{tab_weights_alpha0}. This actually means that weights in $G_W$ are based only on attributive similarity. Clearly, this may lead to dominance of  attributes in the resulting fusion and disappearance of structural connections between nodes with dissimilar attributes. Varying $\alpha$ over the segment $[0,1]$ in (\ref{main_weight}), used by the methods in Table~\ref{tab_weights}, seems more adequate for controlling the impact of both the components. However, tuning $\alpha$ in this case is usually performed {\it manually}. In fact, we are unaware of any general non-manual approaches for tuning $\alpha$ in (\ref{main_weight}), although the need for such approaches has been repeatedly emphasized \cite{Bothorel2015}.

Note that the choice of similarity functions $w_S$ and $w_A$ is usually determined by preferences of the authors of a particular method. The systematic study of how such a choice influences the community detection results is yet to be done.

\begin{table}
	\caption{Weight-based methods with $\alpha=0$ in (\ref{main_weight}).}
	\label{tab_weights_alpha0} 
	\begin{center}
		{\tiny
			\begin{tabular}{|p{1.5cm}|p{4.0cm}|p{1.5cm}|p{1.0cm}|p{1.1cm}|p{1cm}|p{1.3cm}|p{1.5cm}|}
				\hline 
				Method  & Community detection method used and its {\bf input} & Require the number of clusters/ Clusters overlap & Size of datasets used for evaluation & Quality measures & Topology & Datasets used & Compared with \\
				\hline 
				{\bf WNev}~\cite{Neville2003}  & {\bf Weighted graph}\newline MinCut \cite{Karger1993}\newline MajorClust \cite{SteinNiggemann1999}\newline Spectral \cite{ShiMalik2000} & No/No & Small & Accuracy  & Fixed & Synthetic    & --- \\
				\hline 
				{\bf WSte1}~\cite{Steinhaeuser}  & {\bf Weighted graph}\newline Threshold & No/No &  Large & Modularity & Fixed & Phone Network \cite{Madey2007}   &  --- \\
				\hline 
				{\bf WSte2}~\cite{Steinhaeuser2010} & {\bf Similarity matrix (via Weighted graph and random walks)}\newline Hierarchical clustering \cite{Johnson1967,Fred2002} & No/No &  Large & Modularity & Fixed & Phone Network \cite{Madey2007}   &  --- \\
				\hline 
				{\bf WCom1} \cite{Combe2012} & {\bf Weighted graph}\newline Weighted Louvain \cite{Blondel2008} & Yes/No & Small & Accuracy  & Fixed & {\bf DBLP*} &  {\bf WCom2} \cite{Combe2012}\newline {\bf DCom} \cite{Combe2012} \\
				\hline 
				{\bf WCom2} \cite{Combe2012}  & {\bf Distance matrix (via weighted graph)}\newline Hierarchical agglomerative clustering & Yes/No & Small & Accuracy  & Fixed & {\bf DBLP*} &  {\bf WCom1} \cite{Combe2012}\newline {\bf DCom} \cite{Combe2012} \\
				\hline
				{\bf AA-Cluster} \cite{Akbas2017,Akbas2019} & {\bf Node embeddings (via weighted graph)}\newline $k$-medoids  & Yes/No & Small\newline Medium\newline Large & Density\newline Entropy & Fixed & {\bf Political Blogs}\newline  {\bf DBLP*}\newline {\bf Patents*} \newline Synthetic  & {\bf SA-Cluster}~\cite{Zhou2009}\newline {\bf BAGC} \cite{Xu2014}\newline {\bf CPIP} \cite{Liu2015}
				\\
				\hline 
				{\bf PWMA-MILP} \cite{Alinezhad2019}    & {\bf Weighted graph}\newline Linear programming MILP~\cite{Alinezhad2019}  & No/No & Small & RI \newline NMI  & Fixed & {\bf WebKB}  &  --- \\
				\hline 
				{\bf KDComm} \cite{Bhatt2019} & {\bf Weighted graph} \newline Iterative Weighted Louvain & No/No & Small\newline Medium \newline Large & $F$-measure\newline Jaccard measure \newline Rank Entropy measure & Fixed & 	{\bf ego-G+}  \newline {\bf Twitter*} \newline {\bf DBLP*} \cite{Jia2017} \newline Reddit \href{https://archive.org/details/2015_reddit_comments_corpus}{link}  & {\bf CPIP} \cite{Liu2015}\newline {\bf JCDC} \cite{Zhang2016} \newline {\bf UNCut} \cite{Ye2017}\newline {\bf SI} \cite{Newman2015} \\
				\hline 
			\end{tabular}
		}
	\end{center}
\end{table}

\begin{table}
	\caption{Weight-based methods with $\alpha\in[0,1]$ in (\ref{main_weight}).}
	\label{tab_weights} 
	\begin{center}
		{\tiny
			\begin{tabular}{|p{1.25cm}|p{1.5cm}|p{2.0cm}|p{1.5cm}|p{0.8cm}|p{1.2cm}|p{1.2cm}|p{1.7cm}|p{1.7cm}|}
				\hline 
				Method & $\alpha$ in (\ref{main_weight})   & Community detection method used and its {\bf input} & Require the number of clusters/ Clusters overlap & Size of datasets used for evaluation & Quality measures & Topology & Datasets used & Compared with \\
				\hline 
				{\bf WWan} \cite{Wang2010}   & $[0,1]$ in theory\newline $\tfrac{1}{2}$ in experiments & {\bf Edge similarity matrix (via weighted graph)}\newline EdgeCluster \cite{Tang2009}\newline($k$-means variant)  & Yes & Small & NMI  \newline Micro-F1 \newline Macro-F1  & Non-fixed: removing edges & Synthetic \newline BlogCatalog\newline {\bf Delicious$^*$}  &  Non-overlapping co-clustering \cite{Dhillon2003}\\
				\hline 
				{\bf SAC2} \cite{DangViennet2012} & $[0,1]$  & {\bf $k$NN (unweighted) graph (via weighted graph)}\newline (Unweighted) Louvain \cite{Blondel2008} & No/ No & Small\newline Medium & Density\newline Entropy & Non-fixed: removing edges & {\bf Political Blogs}  \newline {\bf Facebook100} \newline {\bf DBLP10K}  & {\bf SAC1} \cite{DangViennet2012}\newline {\bf WSte2} \cite{Steinhaeuser2010}\newline Fast greedy \cite{Clauset2004} for weighted graph \\
				\hline  
				{\bf WCru}~\cite{Cruz2011,CruzBathorelPoulet2012}
				& $[0,1]$ in theory \newline Not specified in experiments & {\bf Weighted graph}\newline Weighted Louvain~\cite{Blondel2008}  & No & Medium & Modularity\newline Intracluster distance & Fixed  & {\bf Twitter$^*$}   & ---\\
				\hline 
				{\bf CODICIL} \cite{Ruan2013} & $[0,1]$ in theory\newline
				$1/2$ in some experiments  & {\bf Weighted graph}\newline Metis \cite{Karypis1998}\newline Markov Clustering~\cite{Satuluri2009} & No  & Small\newline Medium\newline Large & $F$-measure  & Non-fixed: adding and removing edges & {\bf CiteSeer*}\newline {\bf Flickr*}\newline  Wikipedia*  &  {\bf Inc-Cluster} \cite{Zhou2010}\newline  {\bf PCL-DC} \cite{Yang2009} \newline  Link-PLSA-LDA \cite{Nallapati2008}
				\\
				\hline 
				{\bf WMen} \cite{Meng2018} & Not specified & {\bf Weighted graph/Distance matrix for the weighted graph}\newline SLPA \cite{Xie2012}\newline  Weighted Louvain \cite{Blondel2008}\newline K-medoids \cite{Yu2018} & Yes-No/\newline Yes-No & Small\newline Medium & NMI \newline $F$-measure \newline Accuracy   & Fixed &  {\bf Lazega} \newline {\bf Research} \newline {\bf Consult} \newline
				LFR benchmark \cite{Lancichinetti2008}  & {\bf CODICIL}~\cite{Ruan2013}\newline {\bf SA-Cluster}~\cite{Zhou2009} \\  
				\hline 
				{\bf PLCA-MILP} \cite{Alinezhad2019}  & $[0,1]$ & {\bf Weighted graph}\newline Linear programming MILP~\cite{Alinezhad2019}  & No/No & Small & RI \newline NMI  & Non-fixed: adding and removing edges & {\bf WebKB}  &  {\bf SCD} \cite{Li2017}\newline  {\bf ASCD} \cite{Qin2018}\newline {\bf SCI} \cite{Wang2016}\newline  PCL-DC \cite{Yang2009} \newline  Block-LDA \cite{Balasubramanyan2011} \\
				\hline
				{\bf kNN-enhance} \cite{Jia2017} & May be thought as $\alpha=1/2$, $k$NN by attributes  & {\bf Distance matrix (of an edge-augmented graph)} \newline $k$NN\newline $k$-means & No/No & Medium & Accuracy \newline NMI \newline F-Measure \newline  Modularity\newline Entropy & Non-fixed: adding edges & {\bf Cora} \newline {\bf Citeseer}\newline {\bf Sinanet}\newline {\bf PubMed Diabetes}\newline {\bf DBLP*}&  {\bf PCL-DC} \cite{Yang2009}\newline PPL-DC \cite{Yang2010}\newline {\bf PPSB-DC} \cite{Chai2013}\newline {\bf CESNA} \cite{Yang2013}\newline {\bf cohsMix} \cite{Zanghi2010}\newline {\bf BAGC} \cite{Xu2012}\newline {\bf GBAGC} \cite{Xu2014}) \newline {\bf SA-Custer} \cite{Zhou2009}\newline {\bf Inc-Cluster} \cite{Zhou2010}\newline {\bf CODICIL} \cite{Ruan2013}\newline GLFM \cite{Li2011}\\
				\hline 
				{\bf IGC-CSM}  \cite{Nawaz2015} (\href{https://github.com/WNawaz/CSM}{source}) & $[0,1]$ in theory\newline $1/2$ in comparison experiments   & {\bf Distance matrix for the weighted graph} \newline $k$-Medoids & Yes/ No & Medium & Density\newline Entropy & Fixed & {\bf Political Blogs}\newline {\bf DBLP10K}  & {\bf SA-Cluster} \cite{Zhou2009}\newline {\bf SA-Cluster-Opt} \cite{Cheng2011} \\
				\hline 
				{\bf AGPFC} \cite{He2019} & $[0,1]$ in theory, manually tuned in experiments   & Fuzzy equivalent matrix \newline $\lambda$-cut set method & No/Yes & Small\newline Medium & Density\newline Entropy & Fixed & {\bf Political Blogs}\newline {\bf CiteSeer} \newline {\bf Cora} \newline {\bf WebKB}  & {\bf SA-Cluster} \cite{Zhou2009}\newline {\bf BAGC} \cite{Xu2012} \\
				\hline 
				{\bf NMLPA} \cite{Huang2019}  & $1/2$   & {\bf Weighted graph}\newline A multi-label propagation algorithm & Yes/ Yes & Medium & $F1$-score \newline  Jaccard Similarity  & Fixed &  {\bf ego-Facebook}\newline {\bf Flickr*} \cite{Ruan2013} \newline {\bf ego-Twitter}  & {\bf CESNA} \cite{Yang2013}\newline {\bf SCI} \cite{Wang2016} \newline {\bf CDE} \cite{Li2018} \\
				\hline 
			\end{tabular}
		}
	\end{center}
\end{table}

\begin{remark} The weight-based strategy has been applied to more general networks than considered in the present survey. For example,  \cite{Berlingerio2011} and \cite{Papadopoulos2015} use a similar scheme to detect communities in {\it multi-layer} networks. Another example is
SANS \cite{Parimala2015} that works with {\it directed} node-attributed graphs. 
Edge weighting similar to (\ref{main_weight}) is also applied in FocusCO	\cite{Perozzi2014-2}. Although it is not a purely unsupervised community detection method (it requires user's preferences on some of the attributes), it can simultaneously extract local clusters and detect outliers in a node-attributed social network.
\end{remark}
\begin{remark}
Let us mention that there exist community detection methods similar ideologically (attributes $\to$ edge weights $\to$ weighted graph node embeddings\footnote{Low-dimensional continuous vector representations of graph nodes. See also Subsection~\ref{class-ef-embedding}.} $\to$ $k$-means) but preceding to recently proposed {\bf AA-Cluster} \cite{Akbas2017,Akbas2019}. Namely, GraphEncoder~\cite{Tian2014} and  GraRep~\cite{Cao2015}  also first convert a node-attributed graph into a weighted one according to (\ref{main_weight}) with $\alpha=0$ and then find corresponding node embeddings. The embeddings are further fed to $k$-means algorithm to detect communities. However, in opposite to \cite{Akbas2017,Akbas2019}, \cite{Tian2014} and  \cite{Cao2015} mostly focus on node embedding techniques for a weighted graph than on the community detection task. 
\end{remark}

\subsection{Distance-based methods}
\label{class-ef-distance}

\begin{table}[b]
\caption {Distance-based methods.}
\label{tab:distance} 
\begin{center}
	{\tiny
		\begin{tabular}{|p{1.0cm}|p{1.3cm}|p{1.8cm}|p{1.5cm}|p{1.2cm}|p{1.5cm}|p{1.1cm}|p{1.5cm}|p{1.8cm}|}
			\hline 
			Method & $\alpha$ in (\ref{distance_main}) & Community detection method used and its {\bf input} & Require the number of clusters/ Clusters overlap & Size of datasets used for evaluation & Quality measures & Topology & Datasets used & Compared with \\
			\hline  
			{\bf DCom} \cite{Combe2012} & $[0,1]$ & {\bf Distance matrix}\newline Hierarchical agglomerative clustering & Yes/No & Small & Accuracy  & Non-fixed: added edges & {\bf DBLP*} &  {\bf WCom1} \cite{Combe2012}\newline {\bf WCom2} \cite{Combe2012} \\
			\hline
			{\bf DVil} \cite{Villavialaneix2013,Olteanu2013} & $[0,1]$  & {\bf Distance matrix} \newline Self-organizing maps \cite{Villavialaneix2013,Olteanu2013} & No/No & Small\newline Medium & NMI  & Non-fixed: added edges & Synthetic\newline  \href{http://graphcomp.univ-tlse2.fr/}{Medieval Notarial Deeds}  &  --- \\
			\hline  
			{\bf SToC}\cite{Baroni2017} & Maybe thought depending on the values of $d_S$ and $d_A$  & {\bf Distance matrix}\newline $\tau$-close clustering \cite{Baroni2017} & No/No & Medium\newline Large  & Modularity\newline Within-Cluster Sum of Squares & Non-fixed: added edges  & {\bf DBLP10K}\newline DIRECTORS*\newline DIRECTORS-gcc* & {\bf Inc-Cluster} \cite{Zhou2010} \newline {\bf GBAGC} \cite{Xu2014}  \\
			\hline 
			{\bf @NetGA} \cite{Pizzuti2018} & $\alpha\in[0,1]$ in general\newline $\alpha=1/2$ in experiments   & {\bf Distance matrix}\newline Genetic algorithm & No/No & Medium & NMI  & Non-fixed: added edges & Synthetic  &  {\bf SA-Cluster} \cite{Zhou2009}\newline CSPA \cite{Strehl2003,Elhadi2013}\newline {\bf Selection} \cite{Elhadi2013}\\
			\hline 
			{\bf ANCA} \cite{FalihGrozavu2018,Falih2018} & Maybe thought as $\alpha=1/2$    & {\bf Distance matrix}\newline  $k$-means   & Yes/No & Medium & Adjusted Rand Index\newline NMI \newline Density \newline Modularity \newline Conductance \newline Entropy & Fixed & Synthetic  \newline {\bf DBLP10K} \newline 
			Enron email corpus \newline  &  {\bf SA-Cluster} \cite{Zhou2009}\newline {\bf SAC1-SAC2} \cite{DangViennet2012} \newline {\bf IGC-CSM} \cite{Nawaz2015} \newline {\bf WSte1} \cite{Steinhaeuser} \newline {\bf ILouvain} \cite{Combe2015}\\
			\hline  
		\end{tabular}
	}
\end{center}
\end{table}

\begin{figure}
	\centering
	\includegraphics[width=0.9\linewidth]{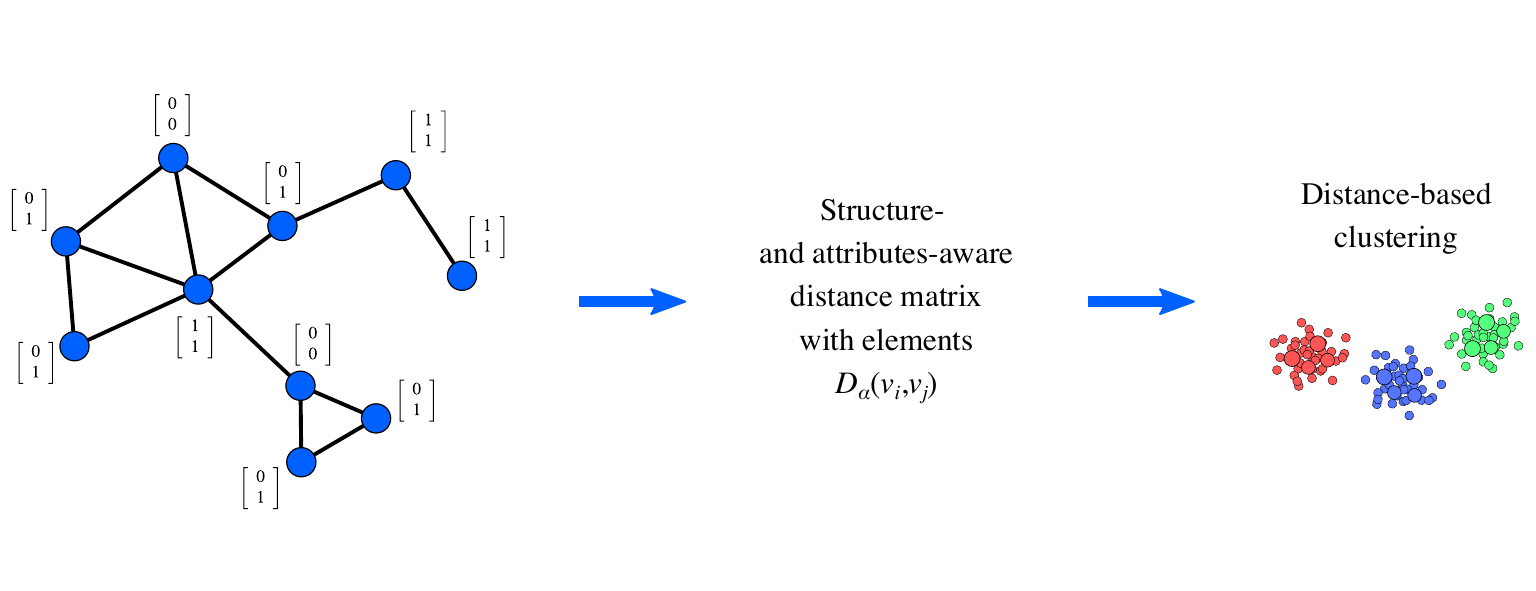}
	\caption{A typical scheme of a distance-based method.}
	\label{fig:distance}
\end{figure}

Methods discussed in the previous subsection aim at representing structure and attributes in a unified graph form suitable for further graph clustering. In opposite, distance-based methods intentionally abandon graph representation in favor of the representation via a distance matrix that contains information about structure and attributes. The distance matrix is usually obtained by fusing  the components by a structure- and attributes-aware distance function, see Figure~\ref{fig:distance}. The matrix can be further fed to distance-based clustering algorithms such as $k$-means and $k$-medoids. Note that the resulting clusters may contain disconnected portions of  initial graph $G$ as the graph structure is removed at the fusion step \cite[Section 3.3]{Akbas2017}.

The distance function fusing  structure and attributes is often of the form
\begin{equation}
\label{distance_main}
D_{\alpha}(v_i,v_j)=\alpha d_S(v_i,v_j)+(1-\alpha)d_A(v_i,v_j),\qquad \alpha\in[0,1],\qquad v_i,v_j\in \mathcal{V},
\end{equation}
where $d_S$ and $d_A$ are {\it structural} and {\it attributive} distance functions, correspondingly. As in the case of (\ref{main_weight}), $\alpha$ in (\ref{distance_main}) controls the balance between the components; how to choose it ``properly'' seems to be an open problem, too. As for the distance functions, it is common to define $d_S(v_i,v_j)$ as the shortest path length between $v_i$ and $v_j$. Possible options for $d_S(v_i,v_j)$ are the Jaccard or Minkowski distances between vectors $A(v_i)$ and $A(v_j)$.

Short descriptions for known distance-based methods are given in Table~\ref{tab:distance}. Note that {\bf ANCA} \cite{FalihGrozavu2018,Falih2018} and {\bf SToC} \cite{Baroni2017} employ a bit different fusion than in (\ref{distance_main}) but nevertheless still deal with certain structure- and attribute-aware distances.

\begin{remark}
There exist distance-based methods for {\it multi-layer} networks. For example, CLAMP \cite{Papadopoulos2017} is an method for clustering networks with heterogeneous attributes and multiple types of edges that uses a distance function similar to (\ref{distance_main}), in a sense.
\end{remark}

\begin{table}
\caption {Node-augmented graph distance-based methods.} \label{tab:sa} 
\begin{center}
	{\tiny
		\begin{tabular}{|p{1.8cm}|p{1.7cm}|p{2.8cm}|p{1.2cm}|p{1cm}|p{1.0cm}|p{1.4cm}|p{2.2cm}|}
			\hline 
			Method & Graph augmentation & Community detection method used and its {\bf input} & Require the number of clusters/ Clusters overlap & Size of datasets used for evaluation & Quality measures &  Datasets used & Compared with\\
			\hline  
			{\bf SA-Cluster} \cite{Zhou2009}\newline {\bf Inc-Cluster} \cite{Zhou2010,Cheng2012}\newline {\bf SA-Cluster-Opt} \cite{Cheng2011}  & New nodes and edges & {\bf Distance matrix (via neighbourhood random walks) }\newline Modified $k$-medoids \cite{Zhou2009} & Yes/No & Small\newline Medium & Density\newline Entropy & {\bf Political Blogs}\newline {\bf DBLP10K}\newline {\bf DBLP84K} &  W-Cluster \cite{Zhou2009} (based on (\ref{distance_main}))\newline {\bf SA-Cluster} \cite{Zhou2009}\newline {\bf Inc-Cluster} \cite{Zhou2010,Cheng2012}\newline {\bf SA-Cluster-Opt} \\
			\hline 
			{\bf SCMAG} \cite{Huang2015} & New nodes and edges & {\bf Distance matrix (via neighbourhood random walks) }\newline Subspace clustering algorithm based on ENCLUS \cite{Cheng1999} & No/Yes & Medium & Density\newline Entropy &  IMDB\newline Arnetminer bibliography$^*$ &  {\bf SA-Custer}  \cite{Zhou2009}\newline GAMer \cite{Gunnemann2014}\\			
			\hline  
			
		\end{tabular}
	}
\end{center}
\end{table}

\subsection{Node-augmented graph-based methods}
\label{class-ef-node-augmented-graph}

\begin{figure}[b]
	\centering
	\includegraphics[width=0.9\linewidth]{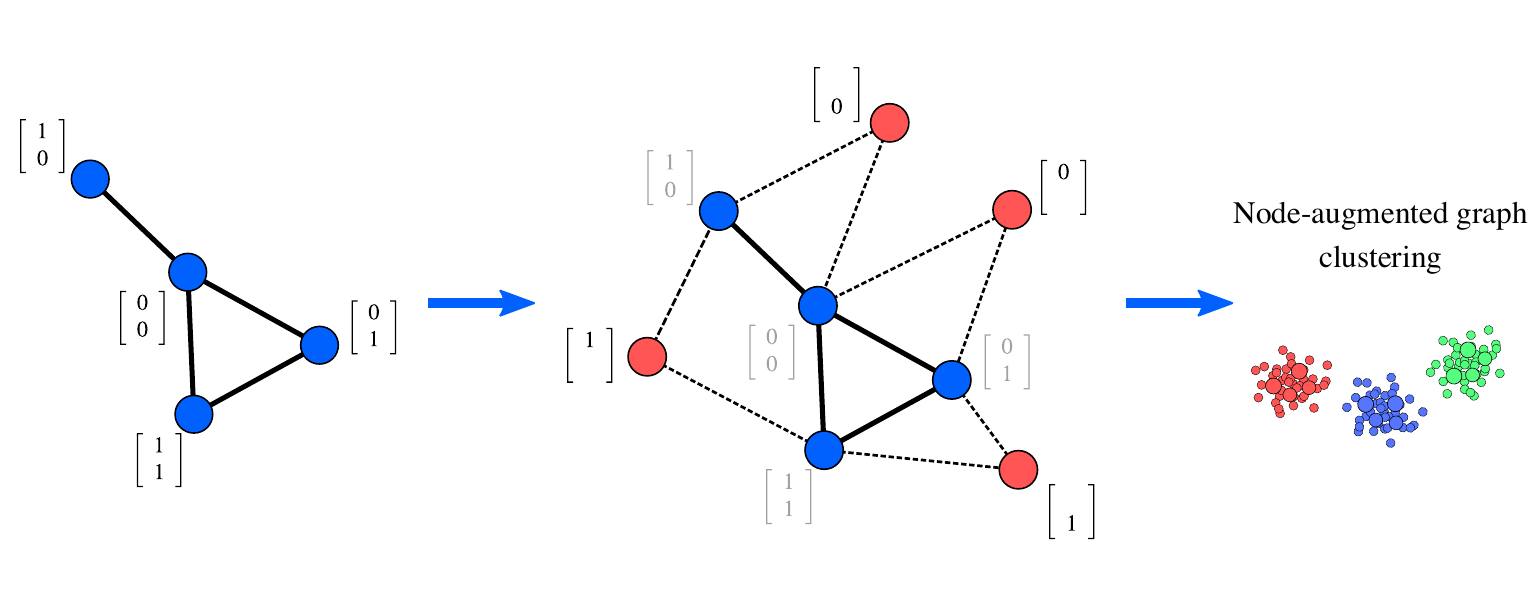}
	\caption{A typical scheme of a node-augmented graph-based method.}
	\label{fig:node-augmented}
\end{figure}

Methods from this subsection (see Table~\ref{tab:sa}) transform  initial node-attributed graph $G$ into another node-augmented graph $\tilde{G}$, with nodes from $\mathcal{V}$ and new {\it attributive} nodes representing attributes, see Figure~\ref{fig:node-augmented}.
To be more precise, suppose that $dom(a_k)=\{s_l\}_{l=1}^{L_k}$, i.e. the domain of $k$th element (attribute) of attribute vector $A$ contains $L_k$ possible values. Then one should create $L_k$ new attribute nodes $\tilde{v}_{k,l}$ corresponding to $l$th value of $k$th attribute. Such a procedure is performed for $k=1,\ldots,d$, where $d=\dim A$. The set $\mathcal{V}_A:=\{\tilde{v}_{k,l}\}$, where $k=1,\ldots,d$ and $l=1,\ldots,L_k$, is then the set of attributive nodes. An edge between structural node $v_i\in \mathcal{V}$ and attributive node $\tilde{v}_{k,l}$ in $\tilde{G}$ exists if $a_k(v_i)=s_l$. Community detection is further performed in $\tilde{G}$. The methods in Table~\ref{tab:sa} propose to apply random walks \cite{Tong2006} to obtain a certain distance matrix for $\tilde{G}$ and further use it in a distance-based clustering algorithm.

Note that the above-mentioned augmentation is not applicable to continuous attributes. What is more, $\tilde{G}$ contains much more nodes end edges than $G$ (especially if $d$ and $L_k$, $k=1,\ldots,d$, are large) and this makes the methods from this subclass rather computationally expensive. 

\subsection{Embedding-based methods}
\label{class-ef-embedding}

\begin{figure}
	\centering
	\includegraphics[width=0.9\linewidth]{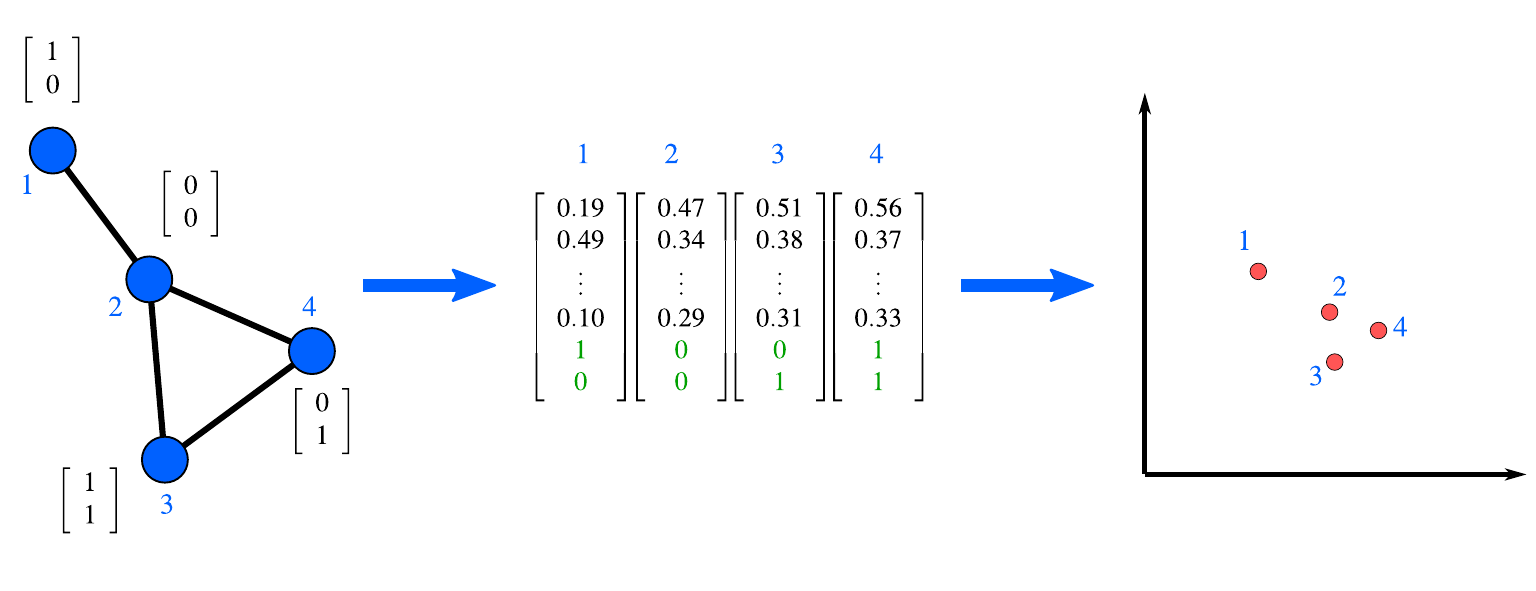}
	\caption{A possible scheme of an embedding-based method (the attribute vectors are here concatenated with the node embeddings and then fed to $k$-means).}
	\label{fig:embeddings}
\end{figure}

As is well-known, a graph as a traditional representation of a network brings several difficulties to network analysis. As mentioned in \cite{Cui2019}, graph algorithms suffer from high computational complexity, low parallelisability and inapplicability of machine learning methods. Novel embedding techniques aim at tackling this by learning {\it node embeddings}, i.e. low-dimensional continuous vector representations for network nodes so that main network information is efficiently encoded \cite{Grover2016,Cui2019,Cai2018}. Roughly speaking, node embedding techniques allow to convert a graph with $n$ nodes into a set of $n$ vectors.

In the context of node-attributed social networks, the objective of node embedding techniques is efficient encoding both structure and attributes \cite{Tang2015,Cao2015,Gao2018}. We are not going to provide general details on the techniques as this has been already done in the surveys \cite{Cui2019,Cai2018}. Let us just mention that having node embeddings (i.e. vectors) at hand allows one to use classical distance-based clustering algorithms such as $k$-means to detect communities, see Figure~\ref{fig:embeddings}.

It turns out that there exists a rich bibliography on node embedding techniques for attributes networks of different type  \cite{Cui2019,Cai2018}. However, not all of them have been applied to the community detection task (the classification task is typically considered).  Taking this into account, we confine ourselves in this survey only to the embedding-based (early fusion) methods that have been used for  community detection in node-attributed social or document networks and compared with other community detection methods. The results are presented in  Table~\ref{tab:emb}.

\begin{table}[b]
\caption {Embedding-based methods.}
\label{tab:emb} 
\begin{center}
	{\tiny
		\begin{tabular}{|p{1cm}|p{1.5cm}|p{2.5cm}|p{1.4cm}|p{1cm}|p{1.9cm}|p{1.5cm}|p{2.3cm}|}
			\hline 
			Method & Embedding technique & Community detection method used and its {\bf input} & Require the number of clusters/ Clusters overlap & Size of datasets used for evaluation & Quality measures &  Datasets used & Compared with\\
			\hline 
			{\bf PLANE} \cite{Le2014} &  A generative model and EM~\cite{Dempster1977}  & {\bf Node embeddings}\newline  $k$-means & Yes/No & Small\newline Medium & Accuracy  &   {\bf Cora*}  &  Relational Topic Model~\cite{Chang2009}+Topic Distributions Embedding~\cite{Iwata2007}\\
			\hline 
			{\bf DANE} \cite{Gao2018} & Autoencoder & {\bf Node Embeddings}\newline $k$-means & Yes/No & Medium & Accuracy  &  {\bf Cora}\newline {\bf Citeseer}\newline {\bf PubMed Diabetes}\newline {\bf Wiki} & Embeddings obtained via TADW \cite{Yang2015}\newline
			LANE \cite{Huang2017b} \newline GAE \cite{Kipf2016} \newline VGAE \cite{Kipf2016}\newline GraphSAGE \cite{Hamilton2017}\\
			\hline 
			{\bf CDE} \cite{Li2018} & Structure embedding matrix & {\bf Structure embedding matrix and attribute matrix}\newline Non-negative matrix factorization & Yes/(Yes/No) & Small\newline Medium & Accuracy \newline NMI \newline Jaccard similarity \newline F1-score  &   {\bf Cora}\newline {\bf Citeseer}\newline {\bf WebKB}\newline {\bf Flickr*}  \newline Philosophers \cite{Hunter2004}\newline {\bf ego-Facebook} & {\bf PCL-DC} \cite{Yang2009}\newline  {\bf Circles} \cite{Leskovec2012}\newline {\bf CESNA} \newline \cite{Yang2013}\newline {\bf SCI} \cite{Wang2016} \\
			\hline 
			{\bf MGAE} \cite{Wang2017} & Autoencoder & {\bf Node embeddings} \newline Spectral clustering & Yes/No & Medium & Accuracy\newline NMI \newline $F$-score \newline Precision \newline Recall \newline Average Entropy \newline Adjusted Rand Index &   {\bf Cora}\newline {\bf CiteSeer} \newline {\bf Wiki} &  {\bf Circles} \cite{Leskovec2012}\newline  RTM \cite{Chang2009} \newline RMSC \cite{Xia2014} \newline Embeddings obtained via TADW \cite{Yang2015} \newline VGAE \cite{Kipf2016}\\
			\hline 
		\end{tabular}
	}
\end{center}
\end{table}

\begin{remark}
Various embedding techniques applicable for community detection in {\it multi-layer} networks are considered e.g. in \cite{Chang2015,Huang2017,Pei2018}.
\end{remark}

\begin{table}[b]
\caption{Pattern mining-based (early fusion) methods.}
 \label{tab:pattern} 
\begin{center}
	{\tiny
\begin{tabular}{|p{1.1cm}|p{0.9cm}|p{2.9cm}|p{2.5cm}|p{2cm}|p{1.5cm}|p{1.2cm}|p{1cm}|}
			\hline 
			Method & Pattern used  & Community detection method used and its {\bf input}  & Require the number of clusters/ Clusters overlap & Size of datasets used for evaluation & Quality measures &  Datasets used & Compared with\\
			\hline 
			{\bf AHMotif} \cite{Li2018Motif}  & Motif & {\bf Structure- and attributes-aware adjacency matrix} \newline Permanence \cite{Chakraborty2014} \newline Affinity Propagation \cite{Frey2007} & Yes/No & Medium & NMI \newline Accuracy  &  {\bf Cora}\newline {\bf WebKB} &  ---\\
			\hline 
		\end{tabular}
	}
\end{center}
\end{table}

\subsection{Pattern mining-based (early fusion) methods}
\label{class-ef-patterns}

Recall that a motif is a pattern of the interconnection occurring in real-world networks at numbers that are significantly higher than those in random
networks \cite{Milo2002}. Motifs are considered as building blocks for complex networks \cite{Milo2002}. We found just one community detection method for node-attributed social networks using such patterns, namely, {\bf AHMotif} \cite{Li2018Motif}, see Table~\ref{tab:pattern}. This method equips structural motifs identified in the network with the so-called homogeneity value based on node attributes involved in the motif. This information is stored in a special adjacency matrix that can be an input to classical community detection algorithms.

\section{Simultaneous fusion methods}
\label{section7}

Recall that simultaneous fusion methods fuse structure and attributes in a joint process with community detection. For this reason, these methods often require special software implementation, in contrast to early and late fusion methods that partially allow one to use existing implementations of classical community detection algorithms.

\begin{table}
	\caption {Methods modifying objective functions of classical clustering algorithms}
	\label{tab_Louvain_NCut} 
	\begin{center}
		{\tiny
			\begin{tabular}{|p{1.2cm}|p{2.8cm}|p{1.8cm}|p{1.5cm}|p{1.6cm}|p{2.0cm}|p{2.4cm}|}
				\hline 
				Method & Modified algorithm  & Require the number of clusters/ Clusters overlap & Size of datasets used for evaluation & Quality measures &  Datasets used & Compared with \\
				\hline 
				{\bf OCru} \cite{Cruz2011}  & Louvain \cite{Blondel2008}\newline  Added attribute Entropy minimization   & No/No & Medium & Modularity\newline Entropy  & {\bf Facebook100}  & --- \\
				\hline 
				{\bf SAC1} \cite{DangViennet2012} & Louvain \cite{Blondel2008}\newline Added attribute similarity maximization  &  No/ No & Small\newline Medium & Density\newline Entropy &  {\bf Political Blogs}\newline {\bf Facebook100} \newline {\bf DBLP10K}  & {\bf SAC2} \cite{DangViennet2012}\newline {\bf WSte2} \cite{Steinhaeuser2010}\newline Fast greedy \cite{Clauset2004} for weighted graph \\
				\hline 
				{\bf ILouvain} \cite{Combe2015} (\href{http://bit.ly/ILouvain}{source})  & Louvain \cite{Blondel2008}\newline Added maximization of attribute-aware Inertia    & No/ No & Small\newline Medium & NMI \newline Accuracy   & DBLP+Microsoft Academic Search$^*$ \newline Synthetic   & ToTeM \cite{Combe2012} \\
				\hline  
				{\bf LAA/LOA} \cite{Asim2017}  & Louvain \cite{Blondel2008}\newline Modularity gain depends on attributes   & No/No & Small & Density\newline Modularity  & \href{https://sites.google.com/site/ucinetsoftware/datasets/covert-networks/londongang}{London gang} \cite{Grund2015} \newline
				\href{https://sites.google.com/site/ucinetsoftware/datasets/covert-networks/italiangangs}{Italy gang} \newline
				\href{http://networkrepository.com/polbooks.php}{Polbooks}
				\newline
				\href{http://networkrepository.com/adjnoun.php}{Adjnoun} \cite{Newman2006}
				\newline
				Football \cite{Girvan2002} & --- \\
				\hline 
				{\bf MAM} \cite{Sanchez2015} (\href{http://ipd.kit.edu/~muellere/mam/}{source})  & Louvain-type algorithm with attribute-aware Modularity+Outlier detection   & No/No & Small\newline Medium \newline Large & F1-score  \newline Attribute-aware Modularity & Synthetic  \newline Disney \cite{Muller2013} \newline DFB \cite{Gunnemann2013} \newline ARXIV \cite{Gunnemann2013} \newline
				IMDB \cite{Gunnemann2013} \newline 
				{\bf DBLP*} \newline
				{\bf Patents*} \newline
				Amazon \cite{Sanchez2013}  & CODA \cite{Gao2010} \\
				\hline 
				{\bf UNCut} \cite{Ye2017} & Normalized Cut\newline Added attributes-aware Unimodality Compactness  & Yes/No & Small\newline Medium & NMI \newline ARI \newline   & Disney \cite{Muller2013} \newline
				DFB \cite{Gunnemann2013} \newline
				ARXIV \cite{Gunnemann2013} \newline
				{\bf Political Blogs} \newline
				4area \cite{Perozzi2014-2} \newline
				{\bf Patents}  & {\bf SA-cluster} \cite{Zhou2009}\newline SSCG \cite{Gunnemann2013}\newline {\bf NNM} \cite{Shiga2007}  \\
				\hline 
				{\bf NetScan} \cite{Ester2006,Ge2008} & An approximation algorithm for the connected $k$-Center optimization problem  (structure and attributes involved) & Yes/Yes & Small\newline Medium & Accuracy & Professors* \newline Synthetic \newline {\bf DBLP*}\newline BioGRID+Spellman
				& --- \\
				\hline 
				{\bf JointClust} \cite{Moser2007}  & An approximation algorithm for the Connected X Clusters problem (structure and attributes involved) & No/No & Medium & Accuracy & {\bf DBLP*} \newline {\bf CiteSeer*} \newline Corel stock photo collection  & --- \\
				
				\hline 
				{\bf SS-Cluster} \cite{Farzi2018} & $k$-Medoid with structure- and attributes-aware objective functions  & Yes/No & Medium & Density \newline Entropy & {\bf Political Blogs}\newline {\bf DBLP10K}  & {\bf SA-cluster} \cite{Zhou2009,Cheng2011}\newline W-cluster \cite{Cheng2011} \newline $k$SNAP \cite{Tian2008} \\
				\hline 
				{\bf Adapt-SA} \cite{Li2019} & Weighted $k$-means for $d$-dimensional representations of structure and attributes   & Yes/No & Medium & Accuracy \newline 
				NMI  \newline F-measure \newline Modularity\newline
				Entropy & Synthetic \newline {\bf WebKB}\newline {\bf Cora}\newline {\bf Political Blogs} \newline {\bf CiteSeer} \newline {\bf DBLP10K}  & {\bf CODICIL} \cite{Ruan2013} \newline {\bf SA-Cluster} \cite{Zhou2006}\newline {\bf Inc-Cluster} \cite{Zhou2010} \newline {\bf PPSB-DC} \cite{Chai2013}\newline {\bf PCL-DC} \cite{Yang2009}\newline  {\bf BAGC} \cite{Xu2012}\\
				\hline 
				{\bf kNAS} \cite{Boobalan2016} & $kNN$ with added Semantic Similarity Score  & Yes/Yes &  Medium & Density\newline Tanimoto Coefficient &  {\bf DBLP*} \newline {\bf Facebook*} \newline {\bf Twitter*} &  {\bf SA-Cluster-Opt} \cite{Cheng2011}\newline {\bf CODICIL} \cite{Ruan2013} \newline NISE \cite{Whang2016}\\
				\hline 
			\end{tabular}
		}
	\end{center}
\end{table}

\subsection{Methods modifying objective functions of classical clustering algorithms}
\label{class-sf-modifying}

Table~\ref{tab_Louvain_NCut} contains\footnote{The authors of {\bf ILouvain} \cite{Combe2015} claim that they compare their method with ToTeM \cite{Combe2012}, ``another community detection method designed for attributed graphs''. However, it seems that there is an inaccuracy with it as we could not find in \cite{Combe2012} any method called ToTeM.} short descriptions of simultaneous fusion methods that modify objective functions of well-known clustering algorithms such as Louvain, Normalized Cut, $k$-means, $k$-medoids and $kNN$. Their main idea is to adapt a classical method (that works, for example, only with network structure originally) for using both structure and attributes in the optimization process. For example, if one wants to modify Louvain \cite{Blondel2008} whose original objective function is structure-aware Modularity, one can include an attributes-aware objective function, say, Entropy in the optimization process. Then Modularity is maximized and Entropy is minimized simultaneously in an iterative process similar to that of Louvain \cite{Blondel2008}.

\subsection{Metaheuristic-based methods}
\label{class-sf-mataheuristic}

Methods in this subclass are rather similar ideologically to those in Subsection~\ref{class-sf-modifying}. However, instead of modifying objective functions and iterative processes of well-known community detection algorithms, they directly apply metaheuristic algorithms (in particular, evolutionary algorithms and tabu search) to find a node-attributed network partition that provides optimal values for certain measures of structural closeness and attribute homogeneity. More precisely, they use  metaheuristics for optimization of a combination of structure- and attributes-aware objective functions, for example, Modularity and Attributes Similarity. Short descriptions of the methods from this subclass are given in Table~\ref{tab_heur}. 

\begin{table}
	\caption {Metaheuristic-based methods.}
	\label{tab_heur} 
	\begin{center}
		{\tiny
			\begin{tabular}{|p{1.2cm}|p{3.7cm}|p{1.5cm}|p{1.0cm}|p{1.4cm}|p{2.5cm}|p{2.2cm}|}
				\hline 
				Method & Metaheuristic optimization algorithm   & Require the number of clusters/ Clusters overlap & Size of datasets used for evaluation & Quality measures &  Datasets used & Compared with \\
				\hline 
				{\bf MOEA-SA} \cite{Li2017-2} & Multiobjective evolutionary algorithm (Modularity and Attribute Similarity are maximized)  & No/No & Small \newline Medium & Density\newline Entropy & {\bf Political Books}\newline {\bf Political Blogs} \newline {\bf Facebook100}\newline {\bf ego-Facebook}  & {\bf SAC1-SAC2} \cite{DangViennet2012} \newline {\bf SA-Cluster} \cite{Zhou2009} \\
				\hline 
				{\bf MOGA-@Net} \cite{Pizzuti2019} & Multiobjective genetic algorithm (optimizing Modularity, Community score, Conductance, attribute similarity)  & No/No & Small\newline Medium & NMI \newline Cumulative NMI \newline Density \newline Entropy & Synthetic \newline {\bf Cora}\newline {\bf Citeseer} \newline {\bf Political books} \newline {\bf Political Blogs} \newline {\bf ego-Facebook} & {\bf SA-cluster} \cite{Zhou2009} \newline {\bf BAGC} \cite{Xu2012}\newline  {\bf OCru} \cite{Cruz2011Entropy}\newline {\bf Selection} \cite{Elhadi2013}\newline HGPA-CSPA \cite{Elhadi2013,Strehl2003}  \\
				\hline
				{\bf JCDC} \cite{Zhang2016} & Tabu search and gradient ascent for a certain structure- and attributes-aware objective function & Yes/No & Small\newline Medium & NMI & Synthetic \newline World trade \cite{Nooy2004} \newline {\bf Lazega}  & CASC \cite{Binkiewicz2017} \newline {\bf CESNA} \cite{Yang2013}\newline {\bf BAGS} \cite{Xu2012} \\
				\hline 
			\end{tabular}
		}
	\end{center}
\end{table}

\subsection{Non-negative matrix factorization-based and matrix compression-based methods}
\label{class-sf-nnmf}
Non-negative matrix factorization (NNMF) is a matrix technique that consists in approximating a non-negative matrix with high rank by a product of non-negative matrices with lower ranks so that the approximation error by means of the Frobenius norm\footnote{A matrix norm defined as the square root of the sum of the absolute squares of matrix  elements.} $F$ is minimal. As is well known, NNMF is able to find clusters in the input data \cite{Lee2001}. 

To be applied to community detection in node-attributed social networks, NNMF requires a proper adaptation to fuse both structure and  attributes. Different versions of such an adaptation have been proposed, see Table~\ref{tab:NMF}. 

To be more formal in describing the corresponding NNMF-based methods, let us introduce additional notation. Let ${\bf S}_{n\times n}$ denote the adjacency matrix for the network structure (as before, $n$ is the number of nodes), ${\bf A}_{n\times d}$ the node attribute matrix for the network attributes ($d$ is the dimension of attribute vector $A$), $N$ the number of required clusters (it is an input in NNMF-based methods), ${\bf U}_{n\times N}$ the cluster membership matrix whose elements indicate the association of nodes with communities, and finally ${\bf V}_{d\times N}$ denotes the cluster membership matrix whose elements indicate the association of the attributes with the communities. In these terms, the aim of NNMF-based methods is to use known matrices  $\bf S$, $\bf A$ and the number of clusters $N$ in order to determine the unknown matrices $\bf U$ and $\bf V$ in an iterative optimization procedure.  For example, {\bf SCI} \cite{Wang2016} models structural closeness as $\min_{{\bf U}\ge 0}\|{\bf S}-{\bf U}{\bf U}^T\|^2_F$ and attribute homogeneity as 
$\min_{{\bf V}\ge 0}\|{\bf U}-{\bf A}{\bf V}\|^2_F$. It is also proposed to select the most relevant attributes for each community by adding an $l_1$ norm sparsity term to each column of matrix
$\bf V$. As a result, one obtains the following optimization problem: 
$$
\min_{{\bf U}\ge 0,\,{\bf V}\ge 0}\left(\alpha_1\|{\bf S}-{\bf U}{\bf U}^T\|^2_F+\|{\bf U}-{\bf A}{\bf V}\|^2_F+\alpha_2 \sum\nolimits_j \|{\bf V}(\cdot,j)\|^2_1\right),
$$
where $\alpha_1>0$ controls the impact of structure and $\alpha_1\ge 0$ the sparsity penalty. This problem is further approximately solved in an iterative process according to Majorization-Minimization framework \cite{Hunter2004}.

\begin{table}
\caption {Non-negative matrix factorization-based and matrix compression-based methods.} \label{tab:NMF} 
\begin{center}
	{\tiny
		\begin{tabular}{|p{1.6cm}|p{1.8cm}|p{1.8cm}|p{1cm}|p{2cm}|p{2.5cm}|p{2.8cm}|}
			\hline 
			Algorithm & Factorization/ compression type   & Require the number of clusters/ Clusters overlap & Size of datasets used for evaluation & Quality measures &  Datasets used & Compared with \\
			\hline 
			{\bf NPei} \cite{Pei2015} & 3-factor NNMF & Yes/Yes & Small\newline Medium & Purity  &  {\bf Twitter} \newline DBLP* &   Relational Topic Model \cite{Chang2009}\\
			\hline 
			{\bf 3NCD} \cite{Nguyen2015} & $2$-factor NNMF  & Yes/Yes & Medium\newline Large & F1-score \newline Jaccard
			similarity &   {\bf ego-Facebook}\newline {\bf ego-Twitter}\newline {\bf ego-G+} &  {\bf CESNA} \cite{Yang2013} \\
			\hline 
			{\bf SCI} \cite{Wang2016} & 2-factor NNMF  & Yes/Yes & Medium & ACC \newline NMI \newline GNMI \newline $F$-measure \newline Jaccard similarity  &   {\bf Citeseer}\newline {\bf Cora}\newline {\bf WebKB}\newline {\bf LastFM} &   {\bf PCL-DC} \cite{Yang2009}\newline {\bf CESNA} \cite{Yang2013}\newline {\bf DCM} \cite{Pool2014}\\
			\hline 
			{\bf JWNMF} \cite{Huang2015} & 2-factor NNMF &  Yes/Yes & Small\newline Medium & Modularity\newline Entropy\newline NMI  &   Amazon Fail \href{https://www.ipd.kit.edu/mitarbeiter/muellere/consub/}{dataset} 
			\newline Disney \href{http://www.perozzi.net/projects/focused-clustering/}{dataset} \newline Enron \href{https://www.ipd.kit.edu/mitarbeiter/muellere/consub/}{dataset} \newline DBLP-4AREA \href{http://www.perozzi.net/projects/focused-clustering/}{dataset}\newline {\bf WebKB} \newline {\bf Citeseer} \newline {\bf Cora}  &  {\bf BAGC} \cite{Xu2012}\newline {\bf PICS} \cite{Akoglu2012}\newline SANS \cite{Parimala2015}\\
			\hline 
			{\bf SCD} \cite{Li2017} & 2- and 3-factor NNMF   & Yes/Yes-No & Small\newline Medium & Accuracy \newline NMI  &  {\bf Twitter}\newline {\bf WebKB} &  {\bf SCI} \cite{Wang2016}\\
			\hline
			{\bf ASCD} \cite{Qin2018} & 2-factor NNMF &Yes/Yes-No & Small\newline Medium & ACC \newline NMI \newline $F$-measure \newline Jaccard similarity  &  {\bf LastFM} \newline {\bf WebKB}\newline {\bf Cora} \newline {\bf Citeseer} \newline {\bf ego-Twitter*} \newline {\bf ego-Facebook*}   &  Block-LDA \cite{Balasubramanyan2011}\newline {\bf PCL-DC} \cite{Yang2009}\newline {\bf SCI} \cite{Wang2016}\newline {\bf CESNA} \cite{Yang2013}\newline {\bf Circles} \cite{Mcauley2014}\\
			\hline
			{\bf CFOND} \cite{Guo2019} & $2$- and $3$-factor NMF &   Yes/(Yes/No) & Medium & Accuracy \newline NMI &   {\bf Cora}\newline {\bf CiteSeer}\newline {\bf PubMed}\newline {\bf Attack}\newline Synthetic
			&  GNMF \cite{Cai2008}\newline DRCC \cite{Gu2009} \newline 
			LP-NMTF \cite{WangNie2011}\newline iTopicModel \cite{Sun2009}\\
			\hline 
			{\bf MVCNMF} \cite{He2017} & $2$-factor NMF & Yes/Yes & Small\newline Medium & Density\newline Entropy &   {\bf Political
				Blogs}\newline {\bf CiteSeer}\newline {\bf Cora}\newline {\bf WebKB} \newline ICDM (DBLP*) &  FCAN \cite{Hu2016}\newline  SACTL \cite{Xu2016} \newline {\bf kNAS} \cite{Boobalan2016}\\
			\hline
			{\bf PICS} \cite{Akoglu2012} (\href{http://www.andrew.cmu.edu/user/lakoglu/pubs.html#pics}{source}) & Matrix compression (finding rectangular blocks)   & No/No & Small\newline Medium & Anecdotal and visual study &   Youtube \cite{Mislove2007}\newline {\bf Twitter}* \newline Phonecall \cite{Eagle2009} \newline Device \cite{Eagle2009} \newline {\bf Political Books} (\href{http://www-personal.umich.edu/~mejn/netdata/}{link}) \newline {\bf Political Blogs} & ---\\
			\hline 
		\end{tabular}
	}
\end{center}
\end{table}

\begin{remark}
An NNMF-based community detection method for {\it multi-layer}  networks is given in \cite{Ito2018}.
\end{remark}

Now we briefly describe {\bf PICS} \cite{Akoglu2012}, the method adopting matrix compression\footnote{The aim of compression methods is to find a
shorter form of describing the same information content.} for community detection in node-attributed social networks. {\bf PICS} uses lossless compression principles from \cite{Grunwald2007} to simultaneously compress the network adjacency matrix $\bf S$ and the attribute matrix $\bf A$. As a result of the compression, certain homogeneous rectangular blocks in the matrices can be determined. Groups of the nodes corresponding to the blocks are considered as communities. One should be aware however that nodes within communities found by {\bf PICS} may not be densely connected due to the definition of a community in \cite{Akoglu2012}. 

\subsection{Pattern mining-based (simultaneous fusion) methods}
\label{class-sf-patterns}

Pattern mining in node-attributed social networks focuses on extraction of patterns, e.g. subsets of specific attributes or connections\footnote{An example of a pattern is a maximal clique \cite{Bothorel2015,Khediri2017}. Recall that  a clique is a subset of nodes in an undirected graph such that every two nodes are adjacent.  A clique is called maximal if there is no other clique that contains it.}, in network structure and attributes \cite{Atzmueller2019}. Among other things, this helps to make sense of a network and to understand how it was formed. In the context of community detection, the extracted patterns are used as building blocks for communities.  

There are many papers devoted to  pattern mining in social networks \cite{Atzmueller2019} but the majority of them do not deal with the task of community detection. The ones relevant to the topic of this survey are presented in Table~\ref{tab:clique}. It is worth mentioning here that it is common for pattern mining-based methods to detect communities not in the whole network but in its part only (e.g. \cite{Pool2014,Atzmueller2016}). 

\begin{table}
\caption {Pattern mining-based (simultaneous fusion) methods.}
\label{tab:clique} 
\begin{center}
	{\tiny
		\begin{tabular}{|p{1.4cm}|p{1.2cm}|p{2.5cm}|p{2cm}|p{2.5cm}|p{2.0cm}|p{1.8cm}|}
			\hline 
			Method & Patterns used  & Require the number of clusters/ Clusters overlap & Size of datasets used for evaluation & Quality measures &  Datasets used & Compared with\\
			\hline 
			{\bf DCM} \cite{Pool2014} (\href{http://www.patternsthatmatter.org/software.php}{source}) & Semantic patterns ({\it queries})  & Yes/Yes & Small\newline Medium & Community score\newline Conductance \newline Intra-cluster density \newline Modularity &   {\bf Delicious}\newline {\bf LastFM}\newline {\bf Flickr} &  --- \\
			\hline 
			{\bf COMODO} \cite{Atzmueller2016} & Semantic patterns  & Yes/Yes & Small\newline Medium & Description complexity\newline Community size &   BibSonomy \cite{Benz2010}\newline {\bf Delicious} \newline {\bf LastFM} &  {\bf DCM} \cite{Pool2014}\\
			\hline 
			{\bf ACDC} \cite{Khediri2017}  & Maximal cliques & Yes/Yes & Medium & Density & {\bf Political Blogs} &   {\bf SA-Cluster} \cite{Zhou2009}\newline {\bf SAC1-SAC2} \cite{DangViennet2012}\\
			\hline
		\end{tabular}
	}
\end{center}
\end{table}

\begin{remark}
ABACUS \cite{Berlingerio2013} detects  communities by extracting patterns in {\it multi-layer} networks.
\end{remark}

\subsection{Probabilistic model-based methods}
\label{class-sf-probabilistic-models}

\begin{table}[b]
	\caption{Probabilistic model-based methods.} \label{tab:statistical} 
	\begin{center}
		{\tiny	\begin{tabular}{|p{1.2cm}|p{2.5cm}|p{1.5cm}|p{1.3cm}|p{1.5cm}|p{3cm}|p{2.5cm}|}
				\hline 
				Method & Model features  & Require the number of clusters/ Clusters overlap & Size of datasets used for evaluation & Quality measures &  Datasets used & Compared with \\
				\hline 
				{\bf PCL-DC} \cite{Yang2009} & Conditional Link Model\newline Discriminative Content model & Yes/No & Medium & NMI \newline Pairwise $F$-measure \newline Modularity\newline Normalized cut &   {\bf Cora}\newline {\bf Siteseer} & PHITS-PLSA  \cite{Cohn2001}\newline LDA-Link-Word \cite{Erosheva2004}\newline  Link-Content-Factorization \cite{Zhu2007}\\
				\hline 
				{\bf CohsMix} \cite{Zanghi2010} & MixNet model \cite{Snijders1997}  & Yes/No & Small & Rand Index  &   Synthetic \newline Exalead.com search engine dataset & Multiple view learning \cite{Zhang2006}\newline Hidden Markov Random Field \cite{Ambroise1997}\\
				\hline 
				{\bf BAGC} \cite{Xu2012}\newline {\bf GBAGC} \cite{Xu2014} & a Bayesian treatment on distribution parameters  & Yes/No & Medium & Modularity\newline Entropy &   {\bf Political Blogs}\newline {\bf DBLP10K}\newline {\bf DBLP84K} &  {\bf Inc-Cluster} \cite{Zhou2010}\newline {\bf PICS} \cite{Akoglu2012} \\
				\hline 
				{\bf VEM-BAGC} \cite{Cao2014} & Based on {\bf BAGC} \cite{Xu2012}  & Yes/No & Medium & Modularity\newline Entropy &    {\bf Political Blogs}\newline Synthetic networks &  {\bf BAGC} \cite{Xu2012}\\
				\hline 
				{\bf PPSB-DC} \cite{Chai2013} & Popularity-productivity stochastic block model and discriminative content model  & Yes/No & Medium & normalized mutual information (NMI)\newline Pairwise
				F measure (PWF)\newline Accuracy &  {\bf Cora} \newline {\bf CiteSeer} \newline {\bf WebKB} &  {\bf PCL-DC} \cite{Yang2009}\newline PPL-DC \cite{Yang2010}\\
				\hline 
			{\bf CESNA} \cite{Yang2013} & A probabilistic generative model assuming communities generate network structure and attributes & No/Yes & Medium\newline Large & Evaluation &  {\bf ego-Facebook}\newline {\bf ego-G+} \newline {\bf ego-Twitter}\newline 
			{\bf Wikipedia* (philosophers)}\newline {\bf Flickr} &  {\bf CODICIL} \cite{Ruan2013}\newline {\bf Circles} \cite{Mcauley2014}\newline Block-LDA \cite{Balasubramanyan2011}\\
				\hline 
				{\bf Circles} \cite{Mcauley2014} & A generative model for friendships in social circles & Yes/Yes & Medium\newline Large & Balanced Error Rate  & {\bf ego-Facebook}\newline {\bf ego-G+} \newline {\bf ego-Twitter} &  Block-LDA \cite{Balasubramanyan2011}\newline Adapted Low-Rank Embedding \cite{Yoshida2010}\\
				\hline 
				{\bf SI} \cite{Newman2015} & A  modified version of a stochastic block model \cite{Holland1983}   & Yes/No & Small\newline Medium & Normalized
				mutual information (NMI) & Synthetic \newline High school friendship network \newline Food web of marine species in the Weddell Sea \newline Harvard Facebook friendship network \newline Malaria HVR 5 and 6 gene
				recombination network &  --- \\
				\hline
				{\bf NEMBP} \cite{HeFeng2017} & A generative model with learning method using a nested EM algorithm with belief propagation & Yes/(Yes/No) & Small \newline Medium & Accuracy \newline NMI \newline GNMI \newline F-score \newline Jaccard &  {\bf WebKB} \newline {\bf ego-Twitter*}\newline
				{\bf ego-Facebook*}\newline {\bf CiteSeer} \newline {\bf Cora}\newline {\bf Wikipedia*}\newline {\bf Pubmed} &  Block-LDA \cite{Balasubramanyan2011}\newline {\bf PCL-DC} \cite{Yang2009} \newline {\bf CESNA} \cite{Yang2013} \newline {\bf DCM} \cite{Pool2014} \newline {\bf SCI} \cite{Wang2016}\\
				\hline
				{\bf NBAGC-FABAGC} \cite{Xu2017} & A nonparametric and asymptotic Bayesian model selection method based on {\bf BAGC} \cite{Xu2012} & No/No & Medium & NMI  \newline Modularity \newline Entropy &  Synthetic \newline {\bf Political Blogs} \newline {\bf DBLP10K} \newline {\bf DBLP84K} &  {\bf PICS} \cite{Akoglu2012}\\
				\hline
			\end{tabular}
		}
	\end{center}
\end{table}

Methods from this subclass probabilistically infer the distribution of community memberships for nodes in a node-attributed social network under the assumption that network structure and attributes are generated according to chosen parametric distributions. Generative and discriminative models are mainly used for the inferring. It is worth mentioning that it is though a non-trivial task to ``properly'' choose a priori distributions for structure and attributes \cite{Akbas2017}.

Short descriptions of the methods from this subclass are given in Table~\ref{tab:statistical}. Pay attention that this table does not contain any method preceding to \cite{Yang2009} and this requires the following additional comments. According to \cite{Yang2009}, several probabilistic model-based clustering methods for node-attributed networks had been proposed before \cite{Yang2009}, for example, in  \cite{Cohn2001,Erosheva2004,Nallapati2008}.  However, they focus on node-attributed {\it document} networks which are out of scope of the present survey. That is why they are non included in Table~\ref{tab:statistical}.

\begin{remark}
TUCM \cite{Sachan2012} proposes a generative Bayesian model for detecting communities in {\it multi-layer} networks where different types of interactions between social actors are possible. 
\end{remark}

\subsection{Dynamical system-based and agent-based methods}
\label{class-sf-dynamics}
Methods from this subclass (see Table~\ref{tab:agent}) treat a node-attributed social network as a dynamic system and assume that its community structure is a consequence of certain interactions among nodes (of course, the attributes are thought to affect the interactions). Some of the methods assume that the interactions occur in an information propagation process, i.e. while information is sent to or received from every node. Others comprehend each node as an autonomous agent and develop a multi-agent system to detect communities. Note that these approaches are rather recent and consider community detection in node-attributed social networks from a new perspective. Furthermore, they seem to be efficient for large networks as can be easily parallelized.

\begin{table}
\caption{Dynamical system-based and agent-based methods.}
\label{tab:agent} 
\begin{center}
	{\tiny
		\begin{tabular}{|p{1.2cm}|p{3.2cm}|p{1.7cm}|p{1.5cm}|p{2cm}|p{2cm}|p{1.8cm}|}
			\hline 
			Method & Description &   Require the number of clusters/ Clusters overlap & Size of datasets used for evaluation & Quality measures &  Datasets used & Compared with \\
			\hline 
			{\bf CPIP}-{\bf CPRW} \cite{Liu2015} & Content (information) propagation models: a linear approximate model
			of influence propagation ({\bf CPIP}) and content propagation
			with the random walk principle ({\bf CPRW}) &   Yes/Yes & Medium & F-score\newline Jaccard Similarity\newline NMI & {\bf CiteSeer}\newline {\bf Cora}\newline {\bf ego-Facebook}\newline {\bf PubMed Diabetes} &  Adamic Adar \cite{Adamic2003} \newline {\bf PCL-DC} \cite{Yang2009} \newline {\bf Circles} \cite{Leskovec2012} \newline {\bf CODICIL} \cite{Ruan2013} \newline {\bf CESNA} \cite{Yang2013}\\
			\hline 
{\bf CAMAS} \cite{Bu2017} & Each node with attributes as an autonomous agent with influence in a cluster-aware multiagent system &  No/Yes & Medium\newline Large & Coverage Rate\newline Normalized Tightness \newline Normalized Homogeneity\newline F1-Score \newline Jaccard\newline Adjusted Rand Index & Synthetic  \newline  {\bf ego-Facebook}\newline {\bf ego-Twitter*} \newline {\bf ego-G+} &  {\bf CESNA} \cite{Yang2013} \newline EDCAR \cite{GunnemannBoden2013}\\
\hline 
{\bf SLA} \cite{Bu2019} & A dynamic cluster formation game  played by all nodes and clusters in a discrete-time dynamical system &   Yes/No & Medium \newline Large & Density \newline Entropy \newline F1-score \newline  &   {\bf Delicious}\newline {\bf LastFM} \newline  {\bf ego-Facebook}\newline {\bf ego-Twitter*} \newline {\bf ego-G+} &  {\bf CESNA} \cite{Yang2013} \newline EDCAR \cite{GunnemannBoden2013} \\
\hline 
		\end{tabular}
	}
\end{center}
\end{table}

\begin{figure}[b]
	\centering
	\includegraphics[width=0.9\linewidth]{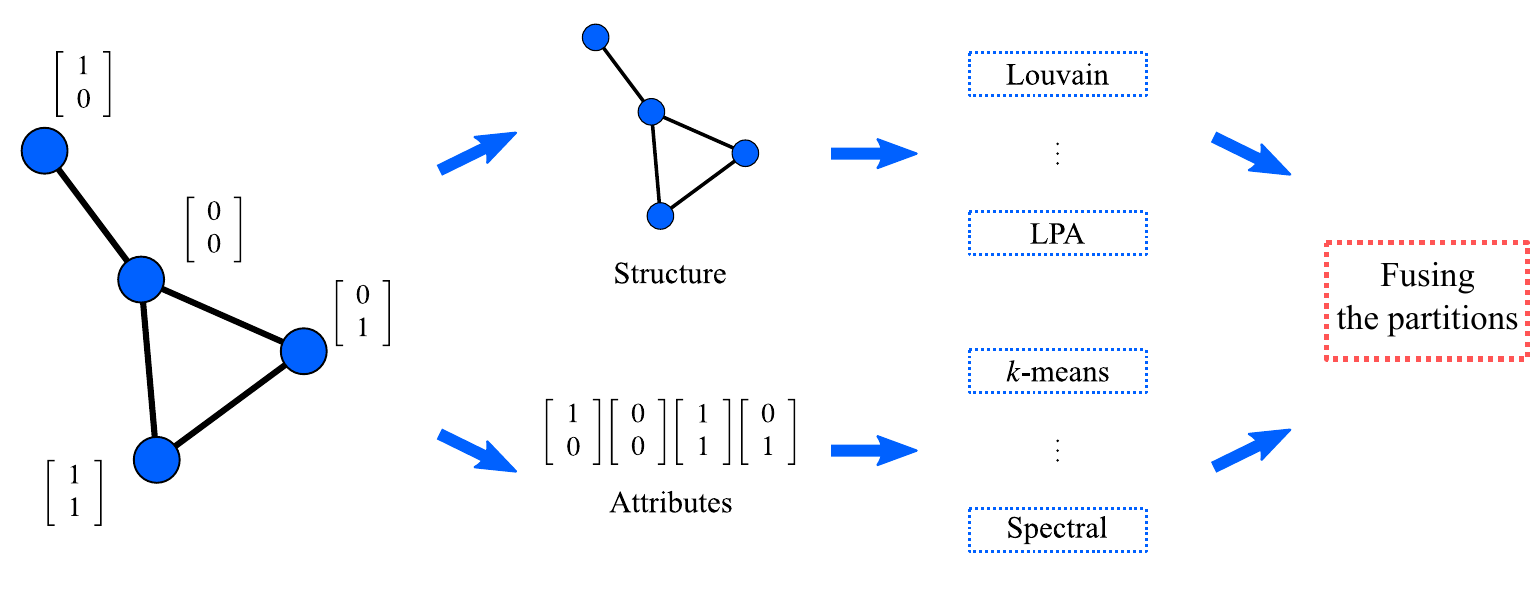}
	\caption{A typical scheme of a late fusion method.}
	\label{fig:late-fusion}
\end{figure}

\section{Late fusion methods}
\label{section8}

Recall that late fusion methods intend to fuse structural and attributive information after the community detection process. More precisely, community detection is first separately performed for structure (e.g. by Louvain \cite{Blondel2008}) and attributes (e.g. by $k$-means \cite{Hartigan1979}). After that, the partitions obtained are fused somehow in order to get the resulting structure- and attributes-aware partition, see Figure~\ref{fig:late-fusion}.

Note that late fusion methods usually allow a researcher/data scientist to use existing implementations of classical community detection and consensus clustering algorithms to get the required partition.

\subsection{Consensus-based methods}
\label{class-lf-consensus}

Given a set (also known as an ensemble) of partitions, the general goal of consensus clustering algorithms is to find a single consolidated partition that aggregates information in the ensemble \cite{Lancichinetti2012,Strehl2003,Tagarelli2017,Tandon2019,Gullo2013}. A recent survey on such methods can be found e.g. in \cite{Boongoen2018}. The idea behind consensus clustering is clearly appropriate for community detection in node-attributed social networks if one has an ensemble of partitions obtained separately (or maybe even jointly) for structure and attributes. Table~\ref{tab:late} contains short descriptions of the methods applying the idea.

\begin{table}[t]
\caption {Consensus-based methods.} \label{tab:late} 
\begin{center}
	{\tiny
		\begin{tabular}{|p{1.2cm}|p{3.0cm}|p{1.9cm}|p{1.5cm}|p{1.5cm}|p{2.3cm}|p{2cm}|}
			\hline 
			Method & Fusing the partitions &   Require the number of clusters/ Clusters overlap & Size of datasets used for evaluation & Quality measures &  Datasets used & Compared with \\
			\hline 
			{\bf LCru} \cite{Cruz2013} & Row-manipulation in the contingency matrix for the partitions &   No/No & Small\newline Medium & ARI \newline Density\newline Entropy   & Facebook$^*$\newline {\bf DBLP10K} &  ---\\
			\hline  
			{\bf Multiplex}\cite{HuangWangg2016} & Multiplex representation scheme (attributes and structure are clustered separately as layers and then combined via consensus \cite{Tepper2015}) &   No/Yes & Medium\newline Large & F1-score &  Synthetic \newline {\bf ego-Twitter} \newline {\bf ego-Facebook} \newline {\bf ego-G+}  &  {\bf CESNA} \cite{Yang2013}\newline {\bf 3NCD} \cite{Nguyen2015} \\
			\hline
			{\bf WCMFA} \cite{Luo2019} & Association matrix with weighting based on structure- and attributes-aware similarity &   Depends on the partitions & Small & Rand index \newline ARI  \newline NMI  &   {\bf Consult}~\cite{Cross2004}\newline 		
			London Gang \cite{Grund2015} \newline
			Montreal Gang \cite{Descormiers2011} &  {\bf WMen} \cite{Meng2018}\\
			\hline 		
		\end{tabular}
	}
\end{center}
\end{table}

\begin{remark}
General-purpose consensus clustering algorithms for {\it multi-layer} networks
are considered in  \cite{Tagarelli2017,Tandon2019,Gullo2013}.
\end{remark}

\subsection{Switch-based methods}
\label{class-lf-switch}

The only method included in this subclass (see Table~\ref{tab:late2}) also deals with partitions obtained separately for structure and attributes but chooses a more ``preferable'' one instead of finding consensus. Namely, {\bf Selection} \cite{Elhadi2013} switches from a structure-based to an attributes-based partition when the former one is {\it ambiguous}. This refers to the case when the so-called {\it estimated mixing parameter} $\mu$ for the structure-based partition is less then a certain experimental value $\mu_{\lim}$ associated with a significant drop in clustering quality on synthetic networks \cite{Lancichinetti2008}. An interested reader can find the precise definitions of $\mu$ and $\mu_{\lim}$ in \cite{Lancichinetti2008,Elhadi2013}.

\begin{table}[t]
\caption {Switch-based methods.} \label{tab:late2} 
\begin{center}
	{\tiny
		\begin{tabular}{|p{1.2cm}|p{3.0cm}|p{1.9cm}|p{1.5cm}|p{1.5cm}|p{2.3cm}|p{2cm}|}
			\hline 
			Method & Fusing the partitions &   Require the number of clusters/ Clusters overlap & Size of datasets used for evaluation & Quality measures &  Datasets used & Compared with \\
			\hline 
			{\bf Selection} \cite{Elhadi2013} & Switching between the partitions &   Depends on the partitions & Medium & NMI \newline Modularity &   Synthetic LFR benchmark \cite{Lancichinetti2008}\newline {\bf DBLP84K} &  {\bf BAGC} \cite{Xu2012}\newline {\bf OCru} \cite{Cruz2011} \newline {\bf SA-Cluster} \cite{Zhou2009} \newline HGPA-CSPA \cite{Strehl2003}\\
			\hline 
		\end{tabular}
	}
\end{center}
\end{table}

\section{Analysis of the overall situation in the field}
\label{section9}

The information collected in Sections~\ref{section6}--\ref{section8} allows us to analyze the overall situation in the field. Particularly, we would like to determine which methods are {\it state-of-the-art}\footnote{According to Cambridge Dictionary, state-of-the-art means ``the best and most modern of its type''.}. This is probably what a researcher/data scientist facing the community detection problem in node-attributed social networks expects from our survey.

\begin{figure}[b]
	\label{fig:graph1}
    \includegraphics[scale=0.95]{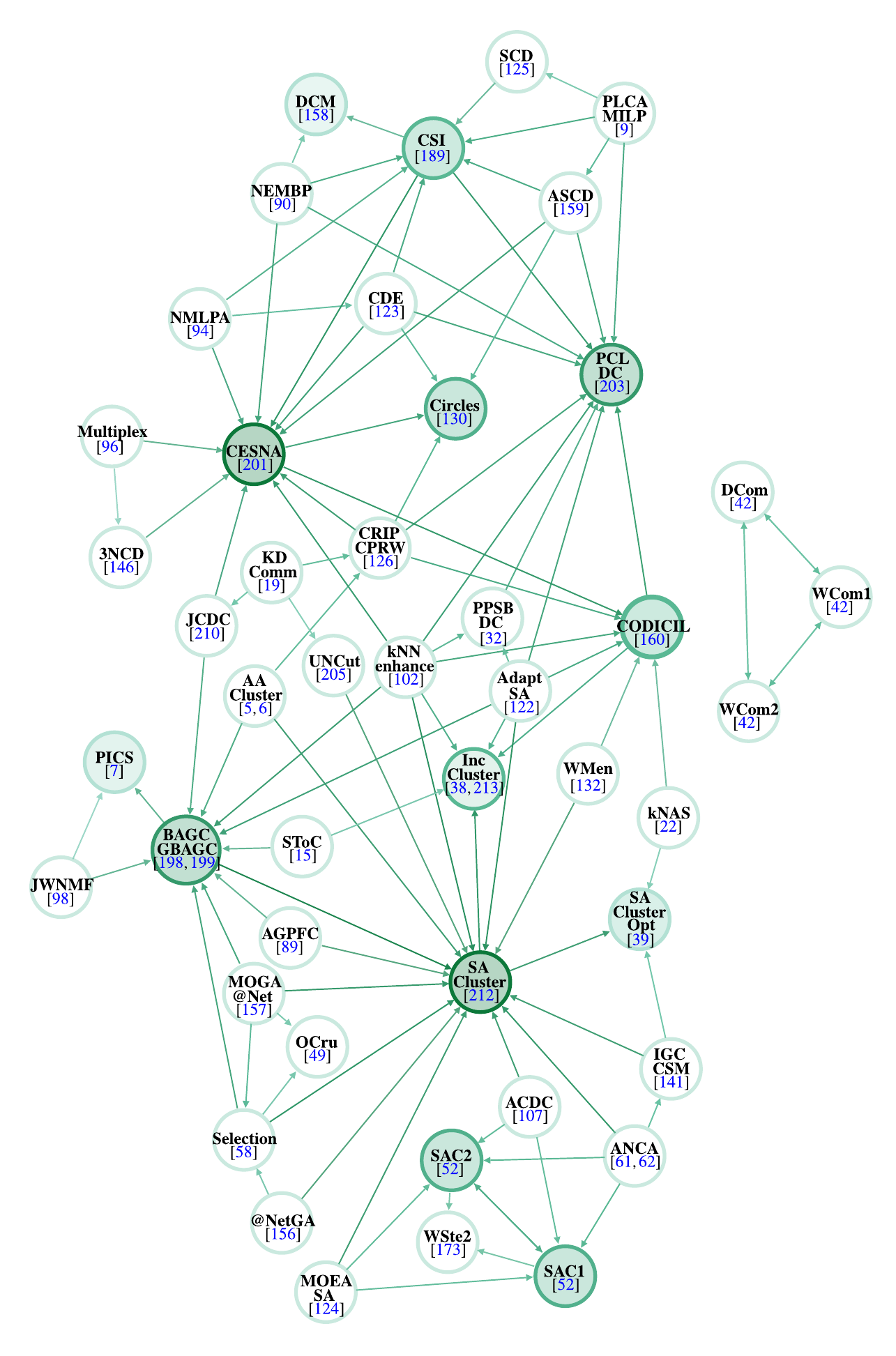}
	\caption{The directed graph of existing method-method comparisons. Nodes (shown only those with degree $\ge 2$) represent the methods classified in the present survey. The most influential methods (nodes with highest PageRank) are filled green so that the darker green means the higher PageRank.}
\end{figure}

We start with observing the directed graph based on the information in Sections~\ref{section6}--\ref{section8} and showing method-method comparisons, see Figure~\ref{fig:graph1}. It requires several comments though. First, there are methods in Sections~\ref{section6}--\ref{section8} that are not compared with others for community detection in node-attributed networks or compared with a few. For this reason, we include in the graph only nodes (representing methods) whose  degree is at least two. This means that there are at least two comparison experiments with each method presented in Figure~\ref{fig:graph1}. Note that 46 methods are shown in the graph of 75 classified in the survey. 
Second, the directed edges in the graph show the existing method-method comparisons. For example, the directed edge from node {\bf CESNA} \cite{Yang2013} to node {\bf CODICIL} \cite{Ruan2013} indicates that the authors of  {\bf CESNA} \cite{Yang2013} compared their method with {\bf CODICIL} \cite{Ruan2013} and showed that {\bf CESNA} \cite{Yang2013} outperforms {\bf CODICIL} \cite{Ruan2013} in some sense (community detection quality, computational efficiency, etc.). This is applied to all edges in the graph.

What is more, we used PageRank to detect the most important or, better to say, {\it most influential methods} in the field. Nodes with the highest PageRank values are filled green in Figure~\ref{fig:graph1} so that the darker green means the higher PageRank. It turns out that the most influential ones are (in the order they discussed in Sections~\ref{section6}--\ref{section8}):
\begin{itemize}
\item weight-based {\bf SAC2}~\cite{DangViennet2012} and {\bf CODICIL}~\cite{Ruan2013} (Subsection~\ref{class-ef-weight}),
\item node-augmented graph-based {\bf SA-Cluster} \cite{Zhou2009}, {\bf Inc-Cluster} \cite{Zhou2010,Cheng2012} and {\bf SA-Cluster-Opt}~\cite{Cheng2011} (Subsection~\ref{class-ef-node-augmented-graph}),
\item {\bf SAC1}~\cite{DangViennet2012} modifying the Louvain objective function   (Subsection~\ref{class-sf-modifying}),
\item NNMF-based {\bf SCI} \cite{Wang2016} and matrix compression-based {\bf PICS} \cite{Akoglu2012} (Subsection~\ref{class-sf-nnmf}),
\item pattern mining-based (simultaneous fusion) {\bf DCM} \cite{Pool2014} (Subsection~\ref{class-sf-patterns}),
\item probabilistic model-based {\bf PCL-DC} \cite{Yang2009}, {\bf BAGC} \cite{Xu2012}, {\bf GBAGC} \cite{Xu2014}, {\bf CESNA} \cite{Yang2013} and {\bf Circles} \cite{Mcauley2014} (Subsection~\ref{class-sf-probabilistic-models}).
\end{itemize}
In our opinion, these methods may be though to be chosen by researchers' community as those determining further developments in the field. Thus we encourage a newcomer in the field  to get familiar with them first to see the main ideas and techniques existing for community detection in node-attributed social networks.
At the same time, we would not consider the most influential methods as state-of-the-art as other methods are shown to outperform them in some sense.

What else the graph in Figure~\ref{fig:graph1} can tell us about? We emphasize that it does not contain 29 methods of those discussed in  Sections~\ref{section6}--\ref{section8} and furthermore it is rather sparse and even disconnected. These points lead to the conclusion that the comparison study of the methods in the field is far from being complete. We would even strengthen the conclusion made by saying that a researcher/data scientist cannot be sure that the method chosen for practical use is preferable to other existing ones, even if there is an edge in Figure~\ref{fig:graph1}. The problem is that the comparison experiments (represented via the edges) are made by different means, e.g. different quality measures, datasets, hyperparameter tuning strategies, etc. Let us discuss it below in more detail.

Suppose for a second that two methods show the same community detection quality for a number of datasets and measures, and their hyperparameters are tuned to provide the best possible results. Then we should think how much time/space each method uses for it. Particularly, we may think of method's computational complexity in terms of the number of vertices $n$, the number of edges $m$ and the attribute dimension $d$ in a node-attributed graph. Such estimates exist for certain methods, particularly for some of the most influential ones. Examples are {\bf CODICIL}~\cite{Ruan2013} with $O(n^2\log n)$, {\bf SA-Cluster} \cite{Zhou2009} with $O(n^3)$ and {\bf CESNA} \cite{Yang2013} with $O(m)$. However, we could not find such estimates for the majority of methods discussed in Sections~\ref{section6}--\ref{section8} as authors often omit such estimation. This makes the overall comparison of methods in terms of computational complexity impossible.

Now let us discuss the hyperparameter tuning problem. It turns out that some authors tune hyperparameters in their methods manually, some just do not consider the problem at all. Another issue is the lack of a general understanding how to determine ``equal impact'' of  structure and attributes on the community detection results. For example, in the weight-based early fusion methods (Subsection~\ref{class-ef-weight}) some authors choose $\alpha=1/2$ in experiments hoping that this provides the equal impact. However, if ones takes into account the different nature of structural and attributive information and the disbalance between associated statistical features, this choice seems questionable in a general situation.
 
Furthermore, authors use different datasets (of various size and nature, see Section~\ref{section5}) and quality measures to test their methods so that a unified comparison of experimental results in different papers cannot be carried out\footnote{A more general observation is that several comparison experiments clearly do not provide generality of conclusions. This however seems to be a hot topic among supporters of scientific research from one side and of empirical one from another side.}. What is more, datasets and software implementations used in comparison experiments are rarely provided by the authors (especially of out-of-date papers) thus making reproducing their results time-consuming or even impossible. Note how few links to source codes are indicated in Sections~\ref{section6}--\ref{section8}.

Let us now discuss the methodology of using quality measures in comparison experiments. There exist two main strategies (see Section~\ref{section5}). Namely, community detection quality can be evaluated (a) by heuristic measures of structural closeness and attributes homogeneity (e.g. Modularity and Entropy) when the dataset under consideration has no ``ground truth'' and (b) by measures estimating the agreement between the detected communities and the ground truth ones (e.g. NMI or ARI).

Let us first discuss Option (a) in terms of a typical weight-based early fusion method from Subsection~\ref{class-ef-weight}. For a given node-attributed graph $G$, suppose that we first convert its attributes $(\mathcal{V},\mathcal{A})$ into attributive graph $G_A$ and further fuse it with the structure $(\mathcal{V},\mathcal{E})$. This results in weighted graph $G_W$ that is thought to contain information about both the structure and the attributes in a unified form. After that, we find communities in $G_W$ that provide, say, the maximum of  Modularity of $G_W$ (e.g. by Weighted Louvain). At the evaluation step we further calculate Modularity of $(\mathcal{V},\mathcal{E})$ and Entropy of $(\mathcal{V},\mathcal{A})$ basing on the partition found. This seems reasonable as far as you do not look at such a methodology critically. Indeed, note that we deal with one quality measure within the optimization process (Modularity of $G_W$) but evaluate community detection performance by other measures (Modularity of $(\mathcal{V},\mathcal{E})$ and Entropy of $(\mathcal{V},\mathcal{A})$). Generally speaking, how can we be sure that optimization of one objective function provides optimal values of other objective functions, if there is no mathematically established connection between them? Of course, this may be simply called a heuristic but it looks more a logical gap, from our point of view. Anyways, we are unaware of any explicit explanation of such a methodology. What is more, one should carefully study how the change of data representation within a method (the change of vectors $\mathcal{A}$ for edge weights in $G_A$ in the above-mentioned example) affects community detection results.

Take into account that a similar discussion is suitable for the majority of methods in Sections~\ref{section6}--\ref{section8} that use quality measures for estimating structural closeness and attributes homogeneity of the detected communities. The exception is the methods that directly optimize the quality measures {\it within the optimization process}. Examples are the simultaneous fusion methods {\bf OCru} \cite{Cruz2011}, 
{\bf SAC1} \cite{DangViennet2012},
{\bf ILouvain} \cite{Combe2015}, 
{\bf UNCut} \cite{Ye2017}, 
{\bf MOEA-SA} \cite{Li2017-2},
{\bf MOGA-@Net} \cite{Pizzuti2019} and 
{\bf JCDC} \cite{Zhang2016}.
Just by construction, they aim at providing   optimal values of the quality measures. One should however take into account the precision of the optimization method applied. 

Now we turn our attention to the methodology of evaluating community detection quality by measures that estimate the agreement between the detected communities and the ground truth ones (Option (b)). In our opinion, such a methodology makes sense for synthetic node-attributed social networks as the way how the communities are formed is known in advance. As for real-world networks, it seems somewhat questionable as ground truth communities may reflect only one point of view on network communities (among many possible). Therefore, expecting that a community detection method would share this point of view seems a bit unsuitable. There are several works on this issue, e.g. \cite{Peele2017,Hric2014,Newman2015}, discussing connections between ground truth, attributes and quality measures in detail, and therefore we refer the interested reader to them.
 
Recall that we started this discussion trying to determine the state-of-the-art methods for community detection in node-attributed social networks. Unfortunately, the above-mentioned facts do not give us a chance to do this. Of course, we could simply list the most recent methods in the field (and then it is enough to check just the time of publication), but this certainly does not meet the requirements imposed on state-of-the-art methods.
 
\section{Conclusions}
\label{section10}

It is shown in the survey that there exist a large amount of methods for community detection in node-attribute social networks based on different ideas. In particular, we gave short descriptions of 75 most relevant methods and mentioned much more those partly related to the topic (Sections~\ref{section6}--\ref{section8}).

We also proposed to divide the methods into the three classes --- early fusion, simultaneous fusion and late fusion ones (Section~\ref{section4}). This classification is based on the moment when network structure and attributes are fused within a method and allows a researcher/data scientist to estimate the ease of method's software implementation. Namely, we concluded that early and late fusion methods can be easier implemented in comparison with simultaneous fusion ones as the former two usually can be combined of several classical community detection algorithms with existing implementations. At a lower lever, we also divided the methods into subclasses of fusion techniques used (Section~\ref{section4}). This allows one to estimate the methodological variety in the field.

Within the classification, we also focused on the experimental part so that one can see which datasets and measures are used for quality evaluation of each method from  Sections~\ref{section5}--\ref{section8}.

The analysis of all the information collected brought us to the unfortunate conclusion that it is impossible now to determine state-of-the-art methods for community detection in node-attributed social networks  (Section~\ref{section9}). This is a result of the presence of the following general problems in the field that we disclosed in the survey:
 \begin{itemize}
  \item the  terminology in the field is rather unstable (Subsection~\ref{subsection 3.2});
  \item there is no generally accepted opinion on the effect of fusing structure and attributes, in particular, on when the fusion is helpful and when not in terms of subclasses of node-attributed social networks (Subsection~\ref{subsection-str-closeness-attr-homo}); 
\item there is no unified methodology for experimental comparison of methods that would include estimation of computational complexity, use of a unified set of datasets and quality measures for evaluation, and justified hyperparameter tuning procedures (Section~\ref{section9});
\item there is no general understanding what is the ``equal impact'' of  structure and attributes on community detection results (Section~\ref{section9}),
     \item as a rule, there is no mathematically established connection between computational processes within a community detection method and the quality measured used for its evaluation (Section~\ref{section9}).
 \end{itemize}
Summarizing, we concluded that the comparison study allowing one to determine the most preferable (in any sense) methods in the field  is far from being complete.

Nevertheless, community detection methods dealing both with network structure and attributes remain a powerful tool in social network analysis and can yield useful insights about a node-attributed social network to a researcher/data scientist. Furthermore, they have wide applications even beyond social networks (Section~\ref{section3}). With respect to these, we believe that the formulation of existing problems in the field done in this survey is the first step in finding solutions to them.

\section*{Competing Interests Statement}

There are no competing interests in publication of this paper.

\section*{Acknowledgements}

The author is very grateful to Klavdiya Bochenina and Timofey Gradov for numerous conversations on the topic and especially to the Anonymous Referee for useful comments that helped to improve the quality of exposition in the survey essentially.

This research is financially supported by Russian Science Foundation, Agreement 17-71-30029 with co-financing of Bank Saint Petersburg, Russia.

{\small
\bibliographystyle{acm}
\bibliography{sample-base}
}
\end{document}